\documentclass[final,3p,times]{elsarticle}

\usepackage{amssymb}
\usepackage{amsthm}
\usepackage{comment}
\usepackage{subfigmat} % enables manipulation of
\journal{Journal of Computational Physics}

\usepackage{amsfonts,amssymb,amsbsy,amsmath}
\usepackage[inline]{enumitem}
\usepackage{physics}
\usepackage{cancel}        % \cancel
\usepackage{listings}      % source code formatting

\usepackage{soul}          % \hl
\usepackage{float}         % float positioning option \begin{figure}[H]
\usepackage{graphicx}      % \includegraphics  
\usepackage{caption}
\usepackage{subcaption}
\usepackage{xcolor}

\usepackage{hyperref} % \autoref

\usepackage{amsthm}
\theoremstyle{definition}

\newcommand{\eqn}[1]{ % numbered equation environment
    \begin{equation}
    \begin{aligned}
        #1
    \end{aligned}
    \end{equation}
}

\newcommand{\dvect}[1]{\bar{#1}}                         % discretized scalar field
\newcommand{\vect}[1]{\boldsymbol{#1}}                         % vector field
\renewcommand{\d}{\mathrm{d}}                                  % total derv

\newcommand{\p}{\partial}                                      % partial derv
\newcommand{\ddd}[1]{\frac{\d}{\d #1}}                         % total derv
\newcommand{\ppp}[1]{\frac{\p}{\p #1}}                         % partial derv
\newcommand{\defeq}{\mathrel{\mathop:}=}                       % define equal to
\newcommand{\mat}[1]{\begin{bmatrix}#1\end{bmatrix}}           % matrix [# & # \\ # & #]
\newcommand{\R}{\mathbb{R}}                                    % Set   of reals
\renewcommand{\bar}[1]{\overline{#1}}

\newcommand{\NN}{\mathrm{NN}}

\DeclareMathOperator*{\argmin}{argmin}

\newcommand{\noteA}[1]{\textcolor{black}{#1}}
\newcommand{\noteB}[1]{\textcolor{black}{#1}}

\usepackage{multirow}

\renewcommand\appendixautorefname[1]{}

\newcommand{\aref}[1]{\hyperref[#1]{\ref*{#1}}}

\usepackage{threeparttable} % table notes
\usepackage{soul} % \st (strike-through text)

\begin{document}

\begin{frontmatter}



\title{
SNF-ROM: Projection-based nonlinear reduced order modeling with smooth neural fields
}


\author[inst1]{Vedant Puri}
\author[inst1]{Aviral Prakash}
\author[inst1]{Levent Burak Kara}
\author[inst1]{Yongjie Jessica Zhang}

\affiliation[inst1]
{organization={Mechanical Engineering, Carnegie Mellon University},
%Department and Organization
            addressline={5000 Forbes Avenue}, 
            city={Pittsburgh},
            postcode={PA 15213}, 
            %state={PA},
            country={USA}
            }

% \author[inst2]{Author Two}
% \author[inst1,inst2]{Author Three}

% \affiliation[inst2]{organization={Department Two},
%             addressline={Address Two}, 
%             city={City Two},
%             postcode={22222}, 
%             state={State Two},
%             country={Country Two}}

\begin{abstract}
% ROM
Reduced order modeling lowers the computational cost of solving PDEs by learning a low-dimensional spatial representation from data and dynamically evolving these representations using manifold projections of the governing equations.
% Linear ROM
The commonly used linear subspace reduced-order models (ROMs) are often suboptimal for problems with a slow decay of Kolmogorov $n$-width, such as advection-dominated fluid flows at high Reynolds numbers.
% ML-ROM
There has been a growing interest in nonlinear ROMs that use state-of-the-art representation learning techniques to accurately capture such phenomena with fewer degrees of freedom.
% our contribution
We propose smooth neural field ROM (SNF-ROM), a nonlinear reduced order modeling framework that combines grid-free reduced representations with Galerkin projection.
% CONTRIB 1
The SNF-ROM architecture constrains the learned ROM trajectories to a smoothly varying path, which proves beneficial in the dynamics evaluation when the reduced manifold is traversed in accordance with the governing PDEs.
% CONTRIB 2
Furthermore, we devise robust regularization schemes to ensure the learned neural fields are smooth and differentiable.
This allows us to compute physics-based dynamics of the reduced system nonintrusively with automatic differentiation and evolve the reduced system with classical time-integrators.
% MISC
SNF-ROM leads to fast offline training as well as enhanced accuracy and stability during the online dynamics evaluation.
\noteA{
Numerical experiments reveal that SNF-ROM is able to accelerate the full-order computation by up to $199\times$.
}
% DEMONSTRATION
We demonstrate the efficacy of SNF-ROM on a range of advection-dominated linear and nonlinear PDE problems where we consistently outperform state-of-the-art ROMs.
\end{abstract}


%\item We present a novel reduced order model based on smooth neural fields (SNF-ROM)
%\item SNF-ROM reduces the order of PDE problems their intrinsic manifold dimension
%\item The online dynamics evaluation is carried out by rapid latent-space traversal
%\item The dynamics evaluation is stable and robust to numerical perturbations by design
%\item We achieves a speed-up of up to $199\times$ over the full-order computation

\begin{keyword}
Reduced order modeling \sep
Neural fields \sep
Physics-based dynamics \sep
Galerkin projection \sep
Scientific machine learning
%%% PACS codes here, in the form: \PACS code \sep code
%%% MSC codes here, in the form: \MSC code \sep code
%%% or \MSC[2008] code \sep code (2000 is the default)
\end{keyword}

\end{frontmatter}


%%%%%%%%%%%%%%%%%%%%%%%%%%%%%%%%%%%%%%%%%%%%%%%%%%%%%%%%%%%%%%%%%%%
% main text

%
%%%%%%%%%%%%%%%%%%%%%%%%%%%%%%%%%%%%%%%%%%%%%%%%%%%%%%%%%%%%%%%%%%%
%
% REVIEWEFRS
% - Miguel Bessa
% - Yungsoo Choi
% - Kookjin Lee
% - Luca Formaggia or Andra Manzoni
% - Jian-Xun Wang
% - Yue Yu
% - Traian Iliescu
%
%%%%%%%%%%%%%%%%%%%%%%%%%%%%%%%%%%%%%%%%%%%%%%%%%%%%%%%%%%%%%%%%%%%
%\newpage
%Journals (in order of preference)
%\begin{enumerate}
%    \item Computer Methods in Applied Mechanics and Engineering
%    \item Journal of Computational Physics
%    \item Journal on Scientific Computing
%    \item SIAM Journal on Scientific Computing
%    \item International Journal for Numerical Methods in Engineering
%    \item IEEE Transactions on Neural Networks and Learning Systems
%    \item Nature Computational Science
%    \item Nature Scientific Reports
%\end{enumerate}

%Conferences
%\begin{enumerate}
%    \item (NeurIPS) Conference on Neural Information Processing Systems
%    \item (ICLR) International Conference on Learning Representatinos
%    \item (ICML) International Conference on Machine Learning
%\end{enumerate}

%%%%%%%%%%%%%%%%%%%%%%%%%%%%%%%%%%%%%%%%%%%%%%%%%%%%%%%%%%%%%%%%%%%
\section{Introduction}
\label{sec:intro}
%%%%%%%%%%%%%%%%%%%%%%%%%%%%%%%%%%%%%%%%%%%%%%%%%%%%%%%%%%%%%%%%%%%
Computational simulations of physical phenomena have become indispensable in scientific modeling and discovery.
However, attaining a reasonable accuracy for practical problems requires resolving a wide range of spatial and temporal scales of motion, leading to large compute and memory costs.
For example, the cost of direct numerical simulations, which attempt to capture all energy-containing length scales in a fluid flow, scales super-linearly with the Reynolds number of the flow, thus limiting the scope of fully resolved simulations to highly simplified problems \cite{hughes2001large, choi_grid-point_2012}.
Therefore, access to high-fidelity numerical partial differential equation (PDE) solutions remains prohibitively expensive for problems requiring repeated model evaluations, such as design optimization \cite{liang2022isogeometric} and uncertainty quantification.
Reduced order models (ROMs) have been studied for the past decade \cite{ahmed_closures_2021} to alleviate the cost of model evaluations for downstream applications like controls and parameter estimation. 
ROMs are hybrid physics and data-based methods that decouple the computation into two stages: an expensive offline stage and a cheap online stage.
In the offline stage, a low-dimensional spatial representation is learned from simulation data by projecting the solution field snapshots onto a low-dimensional manifold that can faithfully capture the relevant features in the dataset.
The online stage then involves evaluating the model at new parametric points by time-evolving the learned spatial representation following the governing PDE system with classical time integrators.
Computational savings are achieved due to the time-evolution of reduced states with much smaller dimensionality than the full order model.
%% DATA-DRIVEN MODELS
%In comparison, purely data-driven models extrapolate poorly outside the training parameter set as learning dynamics from data is not a robust practice \cite{xu_how_2021, shankar_differentiable_2022}.
%As the dynamics in ROMs are equation driven, ROMs extrapolate better to previously unseen parameter points so long as the learned spatial representation can faithfully capture the PDE solution \cite{ahmed_closures_2021}.

% %%%%%%%%%%%%
% \subsection{Background and literature review}
% \label{subsec:litreview}
% %%%%%%%%%%%%
The most commonly used ROMs are linear ROMs, which project the PDE solution onto a linear or affine subspace generated from data.
In the offline stage, a set of global basis functions is computed, typically by applying a variant of proper orthogonal decomposition (POD) \cite{Aubry1988, lucia_reduced-order_2004} to a matrix of solution snapshots. 
These basis functions are computed by projecting the governing PDE problem onto the vector subspace spanned by the POD basis.
%
%Linear ROMs work well when the intrinsic solution space falls into a subspace with a small dimension, i.e., the solution space has a small Kolmogorov $n$-width \cite{GREIF2019216}.
The ability of linear ROMs to compactly capture the physics is described by the
Kolmogorov $n$-width of the data, which is the best possible error one can achieve using a linear approximation with $n$ degrees of freedom,
that is, by projecting to a $n$-dimensional subspace \cite{GREIF2019216}.
Problems that exhibit a fast decay in Kolmogorov $n$-width can be efficiently approximated with smaller ROM representations.
By contrast, problems with a slow decay in Kolmogorov $n$-width, such as many advection-dominated problems, need larger ROM representations or accurate closure models \cite{ahmed_closures_2021, prakash_projection-based_2024} to ensure accuracy while using smaller ROM representations.
Furthermore, most ROMs are intrusive, implying that they rely on access to data-generation source code, which may not be possible when commercial or confidential simulation software is used. These ROMs must rely on specialized methods \cite{ Peherstorfer2016, Prakash2024b} for enabling nonintrusive ROMs.

As an alternative, nonlinear ROMs have been developed using neural networks such as convolutional autoencoders (CAE) for low-dimensional spatial representation
\cite{lee_model_2020, kim_fast_2022, chen_crom_2023, cicci_deep-hyromnet_2022}.
%\cite{lee_model_2020, kim_fast_2022, chen_implicit_nodate, chen_crom_2023, yin_continuous_2023, wan_evolve_2023, chang_licrom_2023, cicci_deep-hyromnet_2022, diaz_fast_2023, cocola_hyper-reduced_2023, wen_reduced-order_2023, pichi_graph_2024}.
CAE-ROMs have been shown to significantly outperform linear ROMs in capturing advection-dominated phenomena with a few degrees of freedom \cite{lee_model_2020, kim_fast_2022}.
Autoencoders produce expressive low-dimensional representations by learning to compress high-dimensional data into a lower-dimensional representation,
where an encoder module maps gridded data to a low-dimensional representation, and a decoder module maps low-dimensional representations back to the grid.

% grid dependent training
Nonlinear ROMs typically use convolutional autoencoder models that limit their applicability to fixed uniform grids.
\noteA{
Furthermore, the evaluation cost for CAE-ROMs scales with the grid size of the full-order computation \cite{lee_model_2020, kim_fast_2022}.
This significantly increases the runtime cost of a ROM,  rendering the method unusable for practical applications.
This limitation has been addressed in \cite{kim_fast_2022} which proposes a shallow masked autoencoder architecture whose inference cost does not scale with the full-order computation.
Another method that addresses this limitation is
the continuous ROM (CROM) \cite{chen_crom_2023, chang_licrom_2023} which employs neural field representations \cite{sitzmann_implicit_2020} that can be queried anywhere in the domain.
Both methods decrease the cost of the online stage by solving the ROM equations at a small set of collocation points. This approach is called hyper-reduction \cite{kim_fast_2022, lauzon_s-opt_2022, chen_crom_2023}.
However, a limitation of both methods is that the encoder module remains grid-dependent and its inference cost scales with the full-order computation.
}
As such, model training requires the problem to have a fixed grid structure, which may be infeasible for adaptive grid simulations and for problems where parts of the data are missing.
% training time
Another major impediment to developing nonlinear ROMs is that training autoencoder networks with stochastic optimization takes much larger compute time than computing a linear ROM representation with POD.
State-of-the-art continuous neural field architectures, such as CROM \cite{chen_crom_2023}, take $O(100,000)$ epochs to train.
As such, training time can be prohibitive even for moderately-sized problems.

% reverse mode AD
%Reverse-mode automatic/algorithmic differentiation, which stores a graph of the forward computation and returns partial derivative values by traversing backward on the graph \cite{SciMLBook}, has been used for this purpose \cite{chen_implicit_nodate}.
%However, this scales poorly when computing higher-order derivatives due to the nesting of computational graphs \cite{chiu_can-pinn_2022}.

Another issue with nonlinear ROMs is that neural field representations are not guaranteed to smoothly interpolate the training dataset \cite{wan_evolve_2023, chen_crom_2023, liu_learning_2022}.
As such, the spatial derivatives of neural field representations are often noisy and may deviate from the true derivative of the underlying signal \cite{chen_crom_2023}.
This can be problematic for the online stage as computing the dynamics of a PDE system entails evaluating partial differential operators like the gradient and the Laplacian of the solution field, as well as the Jacobian of the neural mapping.
Previous nonlinear ROMs have resorted to low-order numerical differentiation on a coarse supplementary mesh for spatial differencing \cite{chen_crom_2023}.
Such workarounds introduce additional memory costs for maintaining a background mesh, and require special treatment for shocks and domain boundaries.
The dynamics evaluation in such cases becomes sensitive to numerical perturbations and the hyperparameters of the spatial differencing scheme due to large approximation errors.
%Besides being inaccurate, such methods introduce new hyperparameters for the spatial differencing scheme that need to be tuned for each problem.

% %%%%%%%%%%%%%%%%
% \subsection{Our contributions}
% \label{subsec:contrib}
% %%%%%%%%%%%%%%%%

In principle, we would like a nonlinear ROM methodology that
(i) can learn from data represented as point-clouds,
(ii) has a low offline stage cost and
(iii) supports fast and robust dynamics evaluation.
In this work, we present smooth neural field ROM (SNF-ROM), a continuous neural field ROM that addresses several of the previously mentioned issues.
% grid-free
SNF-ROM is nonintrusive by construction and eliminates the need for a fixed grid structure in the underlying data and the identification of associated spatial discretization for dynamics evaluation.
% manifold formulation
There are two important features of SNF-ROM:

\begin{enumerate}
    \item\textbf{Constrained manifold formulation:}
    SNF-ROM restricts the reduced trajectories to follow a regular, smoothly varying path.
    This behavior is achieved by directly modeling the ROM state vector as a simple, learnable function of problem parameters and time.
    Our numerical experiments reveal that this feature allows for larger time steps in the dynamics evaluation, where the reduced manifold is traversed in accordance with the governing PDEs.
    
    \item\textbf{Neural field regularization:}
    We formulate a robust network regularization approach encouraging smoothness in the learned neural fields.
    Consequently, the spatial derivatives of SNF representations match the true derivatives of the underlying signal.
    This feature allows us to calculate accurate spatial derivatives with the highly efficient forward mode automatic differentiation (AD) technique.
    Our studies indicate that precisely capturing spatial derivatives is crucial for an accurate dynamics prediction.
\end{enumerate}
The confluence of these two features produces desirable effects on the dynamics evaluation, such as greater accuracy, robustness to hyperparameter choice, and robustness to numerical perturbations.
\noteA{
With a combination of hyper-reduction and use of larger time-steps, SNF-ROM is faster than full-order computation by up to $199\times$.
}
We demonstrate the effectiveness of SNF-ROM using several test cases, such as the scalar transport problem, the Burgers' problem at high Reynolds numbers, and the Kuramoto-Sivashinsky (KS) problem.
Our model implementation and several examples are available on GitHub at
\url{https://github.com/CMU-CBML/NeuralROMs.jl}.

% %%%%%%%%%%%%%%%%%%%
% \subsection{Outline}
% \label{subsec:outline}
% %%%%%%%%%%%%%%%%%%%
The remainder of the paper is organized as follows:
we set up the reduced modeling problem in \autoref{sec:prob};
in \autoref{sec:discr-rom}, we review grid-dependent ROMs and discuss the construction of our SNF-ROM approach in \autoref{sec:manifold-construction};
then \autoref{sec:dynamics} goes over dynamics evaluation in the online stage with Galerkin projection;
numerical examples are presented in \autoref{sec:experiments};
we present conclusions and future work in \autoref{sec:conclusion}.

%%%%%%%%%%%%%%%%%%%%%%%%%%%%%%%%%%%%%%%%%%%%%%%%%%%%%%%%%%%%%%%%%%%
\section{Problem setup}
\label{sec:prob}
%%%%%%%%%%%%%%%%%%%%%%%%%%%%%%%%%%%%%%%%%%%%%%%%%%%%%%%%%%%%%%%%%%%

We seek numerical solutions to the system of PDEs
\begin{gather}
    \label{eqn:prob}
    \ppp{t} \vect{u} = \mathcal{L}(\vect{x}, t, \vect{u}; \vect{\mu}),\,\, \vect{u}: \Omega \times (0, T] \rightarrow \R^m\\
    \vect{u}(\vect{x}, 0; \vect{\mu}) = \vect{u}_0(\vect{x}; \vect{\mu})
\end{gather}
for an $m$-dimensional output field $\vect{u}(\vect{x},t; \vect{\mu}) \in \R^m$ over the spatial domain $\Omega\in\R^d$ with appropriate boundary conditions.
The field $\vect{u}$ may represent a physical quantity like velocity or pressure whose spatio-temporal evolution is dictated by the differential-algebraic operator $\mathcal{L}$.
Let the problem be parameterized by $\vect{\mu}\in\R^p$ using the initial or boundary condition, or $\mathcal{L}$.
% intrinsic manifold dimension
We define the 
\textit{intrinsic manifold dimension} of the problem to be the number of independent variables or degrees of freedom (DoFs) needed to determine the spatial field $\vect{u}(\vect{x}, t; \vect{\mu})$ over $\Omega$ \cite{cicci_deep-hyromnet_2022, fresca_comprehensive_2020, mucke_reduced_2020}.
It is defined as the dimensionality of the smallest parameterization that can approximate the PDE problem.
In \autoref{eqn:prob} the intrinsic manifold dimension is equal to the size of $\vect{\mu}$ plus $1$ (for time)
as $\vect{\mu}$ determines a unique instantiation of the PDE system, and $t$ determines a particular field in the solution trajectory.

%====================================================%
\subsection{Full order model}
%====================================================%

The solution to \autoref{eqn:prob} can be calculated using common discretization methods such as finite difference, finite volume, finite element, or spectral methods.
This computational model is commonly referred to as the full order model (FOM).
Several FOM approaches write the solution to \autoref{eqn:prob} as
\eqn{
    \label{eqn:FOMapprox}
    \vect{u}(\vect{x}, t; \vect{\mu}) = \sum_{i=1}^{N_\text{sp}} \vect{u}_i(t; \vect{\mu}) \phi_i(\vect{x})
}
for $t\in[0,T]$ at some parametric point $\vect{\mu} \in M_\text{FOM} \subset \R^p$.
Here, $\phi_i:\Omega\to\R$ are predetermined spatial basis functions, $\vect{u}_i(t) \in \R^m$ are corresponding time varying coefficients, and
$N_\text{sp}$ is the number of spatial basis functions.
$\phi_i(\vect{x})$ may be locally supported in the case of a finite element method, or globally supported in $\Omega$ if a spectral method is applied.
We collect the FOM DoFs in a single vector
\eqn{
    \dvect{u}(t;\vect{\mu}) \defeq \mat{\vect{u}_1(t;\vect{\mu}) \\ \vdots \\ \vect{u}_{N_\text{sp}}(t; \vect{\mu})} \in \R^{N_\text{FOM}}
}
of size $N_\text{FOM} = m \times N_\text{sp}$.
This is called the \textit{FOM state vector}, as
$\dvect{u}(t; \vect{\mu})$ determines the field $\vect{u}(\cdot, t; \vect{\mu})$.
Therefore, the FOM approximates the PDE solution with a parameterization of size $N_\text{FOM}$.
We define FOM manifold parameterization function
\eqn{
    g_\text{FOM}: \left(\vect{x}, \dvect{u}(t) \right) \to \vect{u}(\vect{x}, t)
}
that maps the FOM state vector to the solution field queried at points $\vect{x} \in \Omega$.
Based on the FOM ansatz in \autoref{eqn:FOMapprox},
the FOM approximation is written with
$g_\text{FOM}$ as
\eqn{
    \label{eqn:FOMmap}
    \vect{u}_\text{FOM}(\vect{x}, t; \vect{\mu})
    \defeq 
    g_\text{FOM}(\vect{x}, \dvect{u}(t; \vect{\mu})) =
    \sum_{i=1}^{N_\text{sp}} \vect{u}_i(t; \vect{\mu}) \phi_i(\vect{x}) % = LBK.
}

We define the FOM projection function
$h_\text{FOM}$ as complementary function to $g_\text{FOM}$ that maps vector fields on $\Omega$ to the corresponding FOM state vectors.
Standard spatial discretizations exhibit a correspondence between the coefficients $\vect{u}_i$ and field values evaluated at mesh points
$X_\text{FOM}=\{\vect{x}_i\}_{i=1}^{N_\text{sp}}\subset\Omega$.
As such, for some $\vect{u}: \Omega \to \R^{m}$, we formulate
\eqn{
    h_\text{FOM}:
    \mat{
        \vdots \\
        \vect{u}(\vect{x}) \\
        \vdots
    }_{\vect{x}\in X_\text{FOM}}
    \to
    \dvect{u}
}
to map from field values on $X_\text{FOM}$ to the FOM state vector $\dvect{u}$,
such that
\eqn{
    \vect{u}(\vect{x})
    \approx
    g_\text{FOM}(
    \vect{x}, \dvect{u}
    ), \, \forall \vect{x}\in X_\text{FOM}.
}
In a linear finite element discretization \cite{baratta_dolfinx_2023}, for example,
$\phi_i$ are Lagrangian interpolants over $X_\text{FOM}$, meaning $\phi_i(\vect{x}_j) = \delta_{ij}$, and
the coefficients $\vect{u}_i(\cdot)$ are equal to the field values at the corresponding points, $\vect{u}(\vect{x}_i, \cdot)$.
When $\phi_i$ are a set of spectral basis functions, the field values and the FOM state are related by a Fourier transform.

With the projection and manifold parameterization functions in place, we commence the time-evolution of the FOM state vector.
The initial FOM state $\dvect{u}(0; \vect{\mu})$ is obtained by applying the projection function to $\vect{u}_0(\vect{x}; \vect{\mu})$ evaluated at $X_\text{FOM}$.
Once $\dvect{u}(0)$ is found, the FOM dynamics are obtained by substituting \autoref{eqn:FOMmap} to \autoref{eqn:prob}.
This returns a system of $N_\text{FOM}$ ODEs describing the evolution of $\dvect{u}(t)$:
\eqn{
    \label{eqn:FOMode}
    \ppp{t} g_\text{FOM}\left(\dvect{u}(t), \vect{x}\right)
    = \mathcal{L}\left(\vect{x}, t, g_\text{FOM}\left(\dvect{u}(t), \vect{x}\right) \right).
}
\autoref{eqn:FOMode} can be solved by time-marching from $t=0$ to $t=T$ over $N_t$ steps given by $T_\text{FOM} = \{t_j\}_{j=1}^{N_t}\subset[0,T]$ following
a time-integration scheme such as Runge-Kutta \cite{dormand_runge-kutta_1986} or Backward Difference Formula \cite{curtiss_integration_1952}.
Once time-evolution is completed, snapshots of saved FOM states, $\{\dvect{u}(t)\}_{t\in T_\text{FOM}}$, can be mapped back to the continuous field $\vect{u}(\vect{x}, t)$ via the FOM parameterization function, $g_\text{FOM}$.
The computational cost of time-evolution of a FOM scales as $N_\text{FOM}$.
As accurate simulations of multiscale phenomena often involve millions to billions of DoFs,
solving a full order PDE problem can be prohibitively expensive for many applications.

%====================================================%
\subsection{Reduced order model}
%====================================================%

A ROM is commonly constructed from the data generated by the FOM evaluated for several time instances at fixed points within the parameter space.
Once ROMs are constructed in the more expensive offline stage, the online stage is typically faster to explore parameters that were not included during the ROM training process.
The size of FOMs is typically much larger than the intrinsic manifold dimension of a PDE problem, making the calculation computationally expensive.
Conversely, ROMs seek spatial representations whose size $N_\text{ROM}$ is close to the intrinsic manifold dimension of the problem.
Employing a reduced spatial representation of size $N_\text{ROM} << N_\text{FOM}$ then leads to smaller ODE systems and, therefore, to a substantial reduction in the cost of the dynamics evaluation.
Therefore, ROMs are typically employed to evaluate many points in a parametric space efficiently.

A ROM representation is defined by a manifold parameterization function $g_\text{ROM}$ that maps ROM state vectors
$\tilde{u}\in\R^{N_\text{ROM}}$ to corresponding field values at query points in $\Omega$.
That is, $g_\text{ROM}$ approximates the mapping
\eqn{
    \vect{u}(\vect{x}, t; \vect{\mu}) \approx
    \vect{u}_\text{ROM}(\vect{x}, t; \vect{\mu}) \defeq
    g_\text{ROM}(\vect{x}, \tilde{u}(t; \vect{\mu})).
}
$g_\text{ROM}$ must be inexpensive to evaluate as it is called multiple times in the online stage.
The set of representable ROM solutions constitutes the ROM manifold
\eqn{
    \label{eqn:manifold}
    \mathcal{M} \defeq
    \Biggl\{
        g_\text{ROM}(\vect{x}, {\tilde{u}})
        \,\big\vert\,
        \forall \tilde{u} \in \R^{N_\text{ROM}}
        ,\,
        \forall \vect{x} \in \Omega
    \Biggr\}.
}
The choice of $g_\text{ROM}$ dictates the quality of the reduced approximation as
a smaller state vector may only be able to express a small class of fields accurately.
We also define the ROM projection function $h_\text{ROM}$ that maps vector fields on $\Omega$ to a corresponding ROM state vector.

Manifold construction, i.e., finding an appropriate $g_\text{ROM}$, involves expensive computations on snapshot data from possibly thousands of FOM evaluations at different $\vect{\mu}$ values.
Specifically, we assume that we have access to snapshot information from FOM simulations performed at parameter values $\vect{\mu} \in M_\text{FOM}$.
Snapshots are taken at times $t\in T_\text{FOM}$, with each snapshot having field values at points $\vect{x}\in X_\text{FOM}$, or equivalently, the FOM configuration vectors at the snapshot times. Our dataset, therefore, is
\eqn{
    \label{eqn:dataset}
    \mathcal{D} =
    \Bigl\{
        \vect{u}(\vect{x}, t; \vect{\mu}) \in \R^m
        \,|\,
        \vect{x}   \in X_\text{FOM}, \, 
        t          \in T_\text{FOM}, \, 
        \vect{\mu} \in M_\text{FOM}
    \Bigl\}.
}

%%%%%%%%%%%%%%%%%%%%%%%%%%%%%%%%%%%%%%%%%%%%%%%%%%%%%%%%%%%%%%%%%%%
\section{Grid dependent manifold representation}
\label{sec:discr-rom}
%%%%%%%%%%%%%%%%%%%%%%%%%%%%%%%%%%%%%%%%%%%%%%%%%%%%%%%%%%%%%%%%%%%

This section discusses reduced representations in linear and nonlinear model order reduction approaches by constructing their respective parameterization and projection functions.
As one typically does not have access to the continuous field $\vect{u}$, previous works
\cite{ahmed_closures_2021,lee_model_2020,kim_fast_2022,diaz_fast_2023}
use a discretized solution field, $\dvect{u}$, for ROMs.
Here,
$g'_\text{ROM}:{\tilde{u}} \to \dvect{u}$,
maps from the ROM state to the FOM state, and composing it with $g_\text{FOM}$ gives
\eqn{
    \label{eqn:ROM_g_FOM}
    g_\text{ROM}(\vect{\vect{x}}, {\tilde{u}})
    =
    g_\text{FOM}(g'_\text{ROM}({\tilde{u}}), \vect{x}).
}
The projection function is obtained similarly by composing
$h'_\text{ROM}: \dvect{u} \to {\tilde{u}}$
with the FOM projection function as
\eqn{
    \label{eqn:ROM_h_FOM}
    h_\text{ROM} = h'_\text{ROM} \circ h_\text{FOM}.
}
This approach leverages the existing grid structure in the FOM data used to compute the reduced space.
This section discusses the construction of $g'_\text{ROM}$ and $h'_\text{ROM}$ for certain linear and nonlinear ROMs.

%====================================================%
\subsection{POD-ROM: Linear subspace representation}
\label{subsec:linearROMs}
%====================================================%
%The PDE solution is then represented as a linear combination of the basis vectors maximally representative of the data matrix.
%The governing PDE is projected onto the vector subspace defined by the basis vectors when the condensed representation is plugged into the governing PDE.
%A small ODE system is then sovled for the scaling coefficients in the online solve.

Linear ROMs such as POD-ROM approximate the discrete FOM solution with a subspace-projection approach, i.e.
$\dvect{u}(t)$ is modeled as a linear (or affine) combination of $N_\text{ROM}$ basis vectors which POD typically computes.
The POD-ROM ansatz,
\eqn{
    \label{eqn:POD-ansatz}
    \dvect{u}(t) \approx
    \dvect{u}_\text{ref} + \sum_{i = 1}^{N_\text{ROM}} \dvect{\psi}_i \tilde{u}_i =
    \dvect{u}_\text{ref} + P_\text{POD} \cdot \tilde{u},
}
is equivalent to projecting the solution field, shifted by $\dvect{u}_\text{ref}\in\R^{N_\text{FOM}}$, onto the column space of
\eqn{
    P_\text{POD} = \mat{\dvect{\psi}_1 & \cdots & \dvect{\psi}_{N_\text{ROM}}} \in \R^{N_\text{FOM} \times N_\text{ROM}}.
}
Then, the ROM state vector $\tilde{u}$ is the coordinate vector of $\dvect{u}$ in the column space of $P_\text{POD}$.
The manifold $\mathcal{M}$ is then a $N_\text{ROM}$ dimensional subspace of $\R^{N_\text{FOM}}$.

The POD matrix $P_\text{POD}$ is computed from a
truncated singular value decomposition (SVD) of the FOM snapshot matrix
\eqn{
    \dvect{U} =
    \mat{
        \cdots &
        \dvect{u}(t; \vect{\mu}) &
        \cdots
    }_{t\in T_\text{FOM}, \, \vect{\mu} \in M_\text{FOM}}
}
whose columns are the FOM state vectors in the dataset $\mathcal{D}$.
The columns of $P_\text{POD}$
are the orthogonal left singular vectors of $\dvect{U}$ corresponding to the largest $N_\text{ROM}$ singular values.
%POD can then be thought of as fitting an $N_\text{ROM}$-dimensional ellipsoid centered at $\dvect{u}_\text{ref}$ to the column vectors of $\dvect{U}$, where each axis of the ellipsoid is represented by a principal component, $\dvect{\psi}_i$.
%The length of each axis corresponds to the variance in that direction.
Based on \autoref{eqn:POD-ansatz}, the POD-ROM manifold parameterization function
\eqn{
    \label{eqn:POD-g}
   g'_\text{POD}(\tilde{u}) = \dvect{u}_\text{ref} + P_{POD} \cdot {\tilde{u}}
}
defines the mapping from the POD coefficients to the FOM space.
% POD-ROM projection
The POD-ROM manifold projection function
\eqn{
    \label{eqn:POD-h}
    h'_\text{POD}(\dvect{u}) = P_{POD}^T \cdot (\dvect{u} - \dvect{u}_\text{ref})
}
then performs the spatial projection of $\dvect{u}\in\R^{N_\text{FOM}}$ onto the column space of $P_\text{POD}$.

%====================================================%
\subsection{CAE-ROM: Nonlinear manifold representation}
\label{subsec:nonlinearROMs}
%====================================================%

%An AE consists of two modules: the encoder module projects the PDE solution on a grid to a reduced bottleneck dimension; the decoder module then recreates the input solution from the reduced vector.
%Encoder-decoder networks, when trained to minimize the error of reconstruction, learn natural low-dimensional representations due to the bottleneck, as illustrated in \autoref{fig:autoencoder}.
%Once the network is trained, the decoder map implicitly defines the nonlinear ROM manifold.
%The decoder maps low-dimensional reduced vectors to the PDE solution on the training grid.

Nonlinear manifold representation methods express the solution field as a nonlinear combination of the ROM state vector.
In the context of discretization dependent ROMs, this means that $g'_\text{ROM}$ is a nonlinear function of ${\tilde{u}}$.
These nonlinear functions often use machine learning (ML) based representations such as autoencoders to represent $g'_\text{ROM}$ \cite{lee_model_2020}.
%Autoencoders, a nonlinear generalization to POD \cite{pichi_graph_2024}, learn to compress high-dimensional data on a fixed grid to a reduced dimension space.
The CAE architecture comprises of two modules: an encoder $e_{\theta_e}$ and a decoder $d_{\theta_d}$ parameterized by weights $\theta_e$ and $\theta_d$ respectively.
The encoder maps high dimensional input data, $\dvect{u}(t)$, to a low-dimensional bottleneck dimension, ${\tilde{u}}(\cdot)\in\R^{N_\text{ROM}}$.
The decoder then recreates the input data from the encoded representation.
When trained to minimize reconstruction errors,
CAEs learn highly expressive low-dimensional neural representations in the bottleneck layer.

We take the bottleneck dimension between the encoder and the decoder to be the space of CAE-ROM state vectors and the learned decoder as the CAE-ROM manifold parameterization
\eqn{
    \label{eqn:CAE-g}
    g'_\text{CAE}: {\tilde{u}} \to d_{\theta_d}({\tilde{u}}).
}
We define the CAE-ROM manifold projection function $h'_\text{CAE}$ to map the FOM state vector $\dvect{u}\in\R^{N_\text{FOM}}$ to the corresponding ROM state vector ${\tilde{u}}\in\R^{N_\text{ROM}}$
such that $h'_\text{CAE}(\dvect{u})$ reproduces $\dvect{u}$ with $g'_\text{CAE}$ to a reasonable accuracy.
That is,
\eqn{
    \dvect{u} \approx g'_\text{CAE} \circ h'_\text{CAE}(\dvect{u}).
}
The implementation of $h'_\text{CAE}$ is detailed in \aref{subsec:appendix-cae-rom-projection}.

The choice of FOM discretization constrains grid-dependent ROM approaches as $g_\text{ROM}$ is restricted to output field values $\vect{u}(\vect{x}_i, t)$ at fixed sensor locations, $X_\text{FOM}$.
This limitation prohibits the development of ROMs from adaptive grid simulation data or from mixing simulation data of different resolutions which is common in multi-fidelity modeling scenarios.
Lastly, in instances where the data is partially corrupted, resulting in incomplete datasets, the manifold parameterization may not be directly possible.

%%%%%%%%%%%%%%%%%%%%%%%%%%%%%%%%%%%%%%%%%%%%%%%%%%%%%%%%%%%%%%%%%%%
\section{SNF-ROM: Offline manifold construction}
\label{sec:manifold-construction}
%%%%%%%%%%%%%%%%%%%%%%%%%%%%%%%%%%%%%%%%%%%%%%%%%%%%%%%%%%%%%%%%%%%

\begin{figure}
    \centering
    \includegraphics[width=\textwidth]{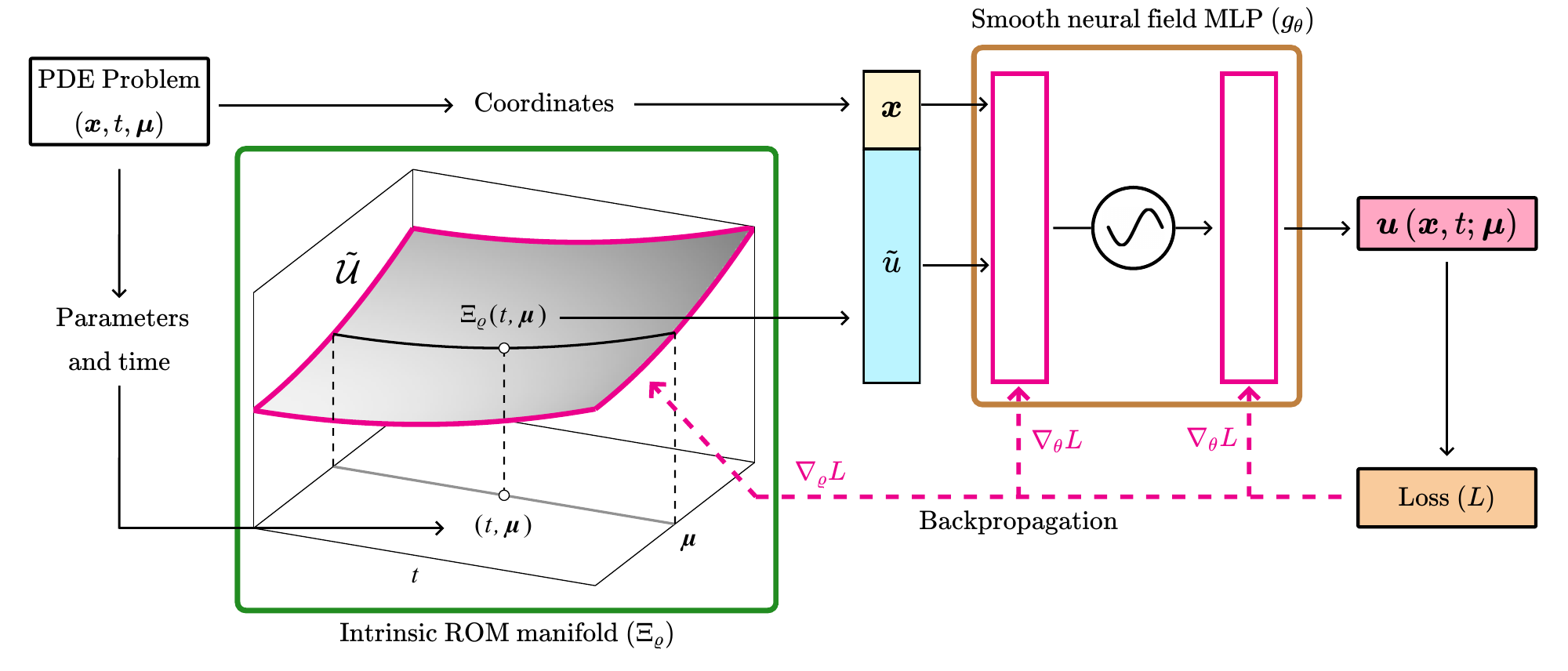}
    \vspace{-2em}
    \caption{
        Offline training procedure for SNF-ROM.
        Given the tuple $(\vect{x}, t, \vect{\mu})$, the forward pass proceeds as follows:
        first, we obtain the ROM state vector corresponding to $t, \vect{\mu}$ on the intrinsic ROM manifold, $\tilde{\mathcal{U}}$, as $\tilde{u} = \Xi_\varrho(t, \vect{\mu})$.
        Directly modeling $\mathcal{\tilde{U}}$ with $\Xi_\varrho$ allows for learning continuous, and smoothly varying ROM state trajectories $\tilde{u}(t; \vect{\mu})$.
        The query point $\vect{x}$ is then concatenated with $\tilde{u}$ and passed to the smooth neural field MLP $g_\theta$ with \textit{sine} activation.
        The output of $g_\theta$ is the model's prediction for $\vect{u}(\vect{x}, t; \vect{\mu}) \in \mathcal{M}$.
        Note that $\tilde{\mathcal{U}}$ is the preimage of $\mathcal{M}$ under $g_\theta$.
        The weights $\theta$ and $\varrho$ are updated with backpropagation to minimize the loss function $L$.
    }
    \label{fig:training-pipeline}
\end{figure}

%Neural fields have recently emerged as useful representations for continuous functions, and have been applied to the development of ROMs \cite{chen_implicit_nodate, chen_crom_2023, chang_licrom_2023} and surrogate models \cite{wan_evolve_2023, yin_continuous_2023, wen_reduced-order_2023, pan_neural_2023}.
%
In this section, we describe the offline training procedure for SNF-ROM.
As illustrated in \autoref{fig:training-pipeline}, our training pipeline operates on point cloud data without the need for any grid structure.
The SNF-ROM manifold parameterization function $g_\theta$ is a neural field,
a multilayer perceptron (MLP) parameterized by learnable weights $\theta$, that maps query points $\vect{x}\in\Omega$ and latent vectors $\tilde{u}\in\R^{N_\text{ROM}}$ directly to field values at that point.
The network $g_\theta$ in this formulation learns common properties of the fields in the dataset $\mathcal{D}$ and embeds them in a low-dimensional space of latent vectors where each latent vector corresponds to a unique field over $\Omega$.
We take the space of latent vectors as the ROM state space and parameterize ${\tilde{u}}$ to continuously vary with time $t$ and problem parameter $\vect{\mu}$ as ${\tilde{u}}(t; \vect{\mu})$.
With this representation of the solution field and state space, we can directly model the continuous field
\eqn{
    \vect{u}_\text{ROM}(\vect{x}, t; \vect{\mu}) =
    g_\theta(\vect{x}, {\tilde{u}}(t; \vect{\mu}))
}
without intermediating through the FOM grid.

In the following subsections, we discuss the formulation of the ROM manifold in \autoref{subsec:manifold-constraint},
and neural field regularization in \autoref{subsec:neural-field-regularization}.
A comprehensive set of implementation details are presented in
\aref{subsec:appendix-snf-rom-projection}, which describes the SNF-ROM projection function $h_\theta$, and in \aref{sec:appendix-snf-rom-train}, which describes the training procedure for SNF-ROM, to supplement the discussion in this section.

%====================================================%
\subsection{Manifold formulation}
\label{subsec:manifold-constraint}
%====================================================%

We cast the manifold construction task as a supervised learning problem to determine SNF-ROM from data.
We seek ROM state vectors
corresponding to the dataset $\mathcal{D}$
\eqn{
    \label{eqn:latentcodeslearn}
    {\tilde{U}} = 
    % https://www.overleaf.com/learn/latex/Brackets_and_Parentheses
    \Bigl\{
        {\tilde{u}}(t; \vect{\mu}) \in \R^{N_\text{ROM}}
        \,\big\vert\,
        t \in T_\text{FOM} ,\,
        \vect{\mu} \in M_\text{FOM}
    \Bigl\}
}
and MLP parameters $\theta$ such that
\eqn{
    \label{eqn:gthetamapping}
    g_\theta(\vect{x}, {\tilde{u}}(t; \vect{\mu}))
    \approx
    \vect{u}(\vect{x}, t; \vect{\mu}),
    \, \forall \vect{x}\in X_\text{FOM},
    \, \forall t \in T_\text{FOM}, 
    \, \forall \vect{\mu} \in M_\text{FOM}.
}
The loss function to be minimized by stochastic optimization is
%We assume Gaussian noise and unit variance on field values $\vect{u}(\cdot)$ to obtain an expression for the data loss:
\eqn{
    \label{eqn:loss_data}
    L_\text{data}(\theta,\, {\tilde{U}};\, \mathcal{D})
    =
    \sum_{\vect{\mu} \in M_\text{FOM}}
    \sum_{t \in T_\text{FOM}}
    \sum_{\vect{x} \in X_\text{FOM}}
    \norm{
        g_{\theta}(\vect{x}, {\tilde{u}}(t; \vect{\mu}))
        - \vect{u}(\vect{x}, t; \vect{\mu})
    }_2^2.
}
Note that a common ROM state vector ${\tilde{u}}(\cdot)$ is learned for pairs $(\vect{x}, \vect{u}(\vect{x}, \cdot))$, and $\theta$ is common for all data pairs.
This ensures that ${\tilde{u}}(\cdot)$ can modulate the decoder to recover the corresponding field $\vect{u}(\vect{x}, \cdot)$ over $\Omega$.

One approach to finding $\theta$ and $\tilde{{U}}$ is to jointly optimize both, following the auto-decode paradigm proposed in \cite{park_deepsdf_2019}.
While minimizing \autoref{eqn:loss_data} is sufficient for learning a neural manifold representation from simulation data, learned nonlinear manifolds exhibit undesirable numerical artifacts in dynamics computation. For example, the learned ROM representations may become disentangled, that is, ROM state vectors corresponding to similar fields are mapped to far away points in $\R^{N_\text{FOM}}$ or vice versa \cite{cha_disentangling_2022}.
An extreme version of this issue is when the learned manifold is disjoint or disconnected, meaning there is no continuous path between ROM state vectors at times $t_1$ and $t_2$.
Another problem with directly optimizing $\tilde{U}$ is that the ROM state vectors are only available for a fixed set of $(t, \, \vect{\mu})$, and any continuous structure in their distribution could be lost.
Our numerical experiments found that these issues can be major impediments in the dynamics evaluation where a continuous trajectory of ROM state vectors is sought.

To resolve above-mentioned issues, we seek a constraint on ${\tilde{U}}$ to enforce the prior that ${\tilde{u}}(t; \vect{\mu})$ are smooth, continuously varying functions of $t$ and $\vect{\mu}$.
This constraint is ensured by modeling the continuous set of ROM state vectors,
\eqn{
    \mathcal{\tilde{U}} \defeq \left\{
        \tilde{u}(t; \vect{\mu}) \in \R^{N_\text{ROM}} \vert\,
        t \in \R_+, \, \vect{\mu} \in \R^p
    \right\},
}
as a learnable function
\eqn{
    \Xi_\varrho\left(t, \vect{\mu} \right) = {\tilde{u}}(t; \vect{\mu})
}
of $t$ and $\vect{\mu}$ that is parameterized by weights $\varrho$.
We refer to $\tilde{\mathcal{U}}$ as the intrinsic ROM manifold as it describes how the ROM state vector varies with the intrinsic coordinates of the PDE problem 
 in \autoref{eqn:prob}.
%
%Similar parameter-to-latent maps have been widely applied in data-driven ROM works for capturing low-dimensional structures in PDE problems \cite{fresca_comprehensive_2020, franco_deep_2022, wen_reduced-order_2023, cho_hypernetwork-based_2023, berman_colora_2024}.
%
As illustrated in \autoref{fig:training-pipeline}, we model the ROM state vector as $\Xi_\varrho(t, \vect{\varrho})$ to enforce continuity and smoothness with respect to both.
Furthermore, we design $\Xi_\varrho$ as a simple, smoothly varying function (see \aref{sec:appendix-snf-rom-train} for details).
This constraint helps to organize the ROM state vectors better, makes them more interpretable, and facilitates meaningful interpolations between data points.
These characteristics are especially helpful in the dynamics evaluation of the reduced system when we traverse the reduced manifold guided by the dynamics of the governing PDE.
Our numerical experiments in \autoref{sec:experiments} indicate that smooth reduced-state dynamics support numerical stability in the dynamic evaluation of the ROM while using larger time steps.

We reformulate the offline stage learning problem to approximate the mapping
\eqn{
    \vect{u}(\vect{x}, t; \vect{\mu}) \approx g_\theta\left(\vect{x}, \Xi_\varrho(t, \vect{\mu}) \right).
}
Note that $\tilde{\mathcal{U}}$ is the preimage of the ROM manifold $\mathcal{M}$ under $g_\theta$.
We simultaneously learn both by jointly optimizing $\varrho$ and $\theta$.
We reformulate the data loss to reflect the same,
\eqn{
    \label{eqn:loss_data-1}
    \bar{L}_\text{data}(\theta, \varrho; \mathcal{D})
    =
    \sum_{\vect{\mu} \in M_\text{FOM}}
    \sum_{t \in T_\text{FOM}}
    \sum_{\vect{x} \in X_\text{FOM}}
    \norm{
        g_{\theta}(\vect{x}, \Xi_\varrho(t; \vect{\mu}))
        - \vect{u}(\vect{x}, t; \vect{\mu})
    }_2^2.
}

%====================================================%
\subsection{Neural field regularization}
\label{subsec:neural-field-regularization}
%====================================================%

When computing the dynamics of the reduced system, the right-hand-side (RHS) term in \autoref{eqn:prob}, $\mathcal{L}(\vect{x}, t, \vect{u})$, must be calculated where the solution field $\vect{u}$ is given by the SNF representation $g_\theta(\vect{x}, {\tilde{u}})$.
This computation involves computing spatial derivatives of neural field representations, which is nontrivial as such representations are not guaranteed to interpolate the training dataset smoothly \cite{liu_learning_2022, wang_piratenets_2024, chetan_accurate_2023}.
The key problem is that neural field ROMs are trained to approximate the signal itself;
there is no guarantee on the quality of the approximation of signal derivatives \cite{chetan_accurate_2023}.
As such, partial derivatives of neural field ROMs are riddled with numerical artifacts, making them unusable in downstream applications such as time-evolution.

In this context, we construct smooth neural fields, which are essentially continuous neural fields with inherent smoothness.
This smoothness is ensured by dampening high-frequency modes by applying regularization during training.
Exact spatial derivatives of the neural field can then be efficiently computed with automatic differentiation.
Regularizing the network smoothens its output in relation to all inputs.
As a consequence, the network's Jacobian with respect to the ROM state vector
$\dfrac{\p}{\p \tilde{u}}g_\theta(\vect{x}, \tilde{u})$ becomes a smoothly varying function.
This feature is extremely advantageous as a smooth Jacobian permits larger time steps in the dynamics evaluation \cite{berman_colora_2024}.
%Regularization also improves network generalization and convergence by reducing the degree to which the model overfits the training dataset \cite{power_grokking_2022, ying_overview_2019}.
Our approaches are based solely on network parameters and can be implemented in a few lines of code with minimal computational overhead.

This section introduces two regularization techniques designed to encourage smooth neural representations.
The first approach is principled in minimizing the Lipschitz constant of an MLP function;
the second is derived by analyzing the derivative expressions of an MLP.
We preface the regularization schemes with the description of a canonical MLP.
Let $\NN_\theta:\R \to \R$ be an MLP with scalar inputs and scalar outputs without loss of generality.
The MLP is composed of $L$ affine layers, $Z_l$, parameterized by weight matrices $W_l$ and bias vectors $b_l$ for $l\in\{1, \cdots, L\}$.
Each layer performs the following affine transformation to input $z$,
\eqn{
    Z_l(z) = W_l \cdot z + b_l.
}
The set of trainable parameters in the network are
\eqn{
    \theta = \Bigl\{W_1, b_1, \ldots, W_L, b_L \Bigr\}.
}
A pointwise activation function, $\sigma$, is applied to the output of all but the last layer.
The action of the neural network on input $z$ is then defined as
\eqn{
    \label{eqn:NN}
    \NN_\theta(z) =
    Z_{L  } \circ \sigma \circ
    Z_{L-1} \circ \sigma \circ
    \cdots
    \circ \sigma \circ Z_2 \circ
    \sigma \circ Z_1
    \left( z \right).
}

%=======================%
\subsubsection{SNFL-ROM: Smooth neural field ROM with Lipschitz regularization}
\label{subsubsec:lipschitz}
%=======================%

As indicated in \cite{liu_learning_2022}, smoothness can be encouraged by minimizing the Lipschitz constant of the network.
A function $f$ is called \textit{Lipschitz continuous} if there exists a constant $c \geq 0$ such that
\eqn{
    \underbrace{\norm{f(x_2) - f(x_1)}_p}_\text{change in output}
    \leq c \, \underbrace{\norm{x_2 - x_1}_p}_\text{change in input}
}
for all possible $x_1, x_2$ under a $p$-norm of choice \cite{liu_learning_2022}.
The parameter $c$ is called the Lipschitz constant of function $f$.
The Lipschitz constant acts as the upper bound to the frequency of noise in a function.
As noisy functions necessarily have large Lipschitz constants, minimizing the Lipschitz constant of an MLP during training could encourage smoothness in the neural representation.
Unfortunately, one does not have direct access to the Lipschitz constant of an MLP; we instead minimize its Lipschitz bound,
\eqn{
    \label{eqn:lipschitz-bound}
    \bar{c}(\theta) = \prod_{l=1}^L \norm{W_l}_p,
}
the upper bound to the Lipschitz constant of the MLP (\autoref{eqn:NN}) when $\sigma$ has Lipschitz constant of unity \cite{liu_learning_2022}.
We formulate a loss function that penalizes the Lipschitz bound of the MLP,
\eqn{
    \label{eqn:loss-lipschitz}
    L_\text{Lipschitz}(\theta) = \alpha \bar{c}(\theta),
}
where the magnitude of regularization is modulated by the hyperparameter $\alpha$.
We set $p=\infty$ and minimize the infinity norm of the weight matrices in the network.
In our experiments of \autoref{sec:experiments}, we verify that this approach produces smooth neural network representations that illustrate a marked reduction in noise compared to other works.
We write the learning problem for smooth neural field ROM with Lipschitz regularization (SNFL-ROM) as
\eqn{
    \label{eqn:learning-prob-lipschitz-reg}
    %\theta, \varrho =
    \argmin_{\theta, \varrho}
    \left\{
        \bar{L}_\text{data}(\theta, \varrho; \mathcal{D}) + L_\text{Lipschitz}(\theta)
    \right\}
    .
}

%=======================%
\subsubsection{SNFW-ROM: Smooth neural field ROM with weight regularization}
\label{subsubsec:weight}
%=======================%

Our second approach is to penalize high-frequency components directly in the expression of the derivative of an MLP.
Let $z_0 \defeq z$, the input to the MLP, and let the output of $Z_l$ be denoted as $z_l$.
Following \cite{wang_piratenets_2024}, the expression for the first derivative is
\eqn{
    \label{eqn:NN-derv}
    \ddd{z} \NN_\theta(z) =
    \left(
        \prod_{l=2}^{L} Z'_{l}(\sigma(z_{l-1}))
        \cdot
        \mbox{diag}(\sigma'(z_{l-1}))
    \right)
    \cdot Z'_{1}(z)
    =
    \left(
        \prod_{l=2}^{L} W_{l}
        \cdot
        \mbox{diag}(\sigma'(z_{l-1}))
    \right)
    \cdot W_1
}
where
$\mbox{diag}(\cdot)$ is a diagonal matrix whose non-diagonal entries are all zero.
We seek to mitigate high-frequency features in \autoref{eqn:NN-derv} and, as such, restrict our attention to the diagonal matrices; the weight matrices in \autoref{eqn:NN-derv} are constant and do not vary with the input.
In this paper, we choose sinusoidal activation functions, that is, $\sigma(z) = \sin(z)$, for their ability to represent complex signals with relative ease \cite{sitzmann_implicit_2020}.
Thus,
\eqn{
    \label{eqn:shallow-NN-derv}
    \sigma'(z_{l-1}) &= \cos(W_{l} \cdot z_{l-1} + b_{l}).
}
In this expression, $W_l$ modulates the frequency and $b_l$ modulates the phase shift of the \textit{cosine}.
The derivative expression would thus be highly oscillatory when the entries in the weight matrices are large, implying that directly smoothing the weight matrices can help remove high-frequency noise in the derivative fields.
%Note that the terms outside the \textit{cosine} do not cause high-frequency noise.
As the weight entries are initialized with zero mean, it is sufficient to penalize their magnitude through the additional loss term,
\eqn{
    \label{eqn:weight-decay}
    L_\text{Weight}(\theta) &=
    \frac{\gamma}{2} \sum_{l=1}^{L}
    \sum_{i=1}^{d_{l}}
    \sum_{j=1}^{d_{l-1}}
    \norm{W_l^{ij}}_2^2,
}
where $W^{ij}$ is the $(i, j)$-th entry of the $W$ matrix.
We write the learning problem for a smooth neural field ROM with weight regularization (SNFW-ROM) as
\eqn{
    \label{eqn:learning-prob-weight-reg}
    %\theta, \varrho =
    \argmin_{\theta, \varrho}
    \left\{
    \bar{L}_\text{data}(\theta, \varrho; \mathcal{D}) + L_\text{Weight}(\theta)
    \right\}
    .
}
Compared to $L_2$ regularization, which penalizes both weights and biases, our weight regularization scheme leads to faster convergence and comparably smooth models.
We believe this is because $L_2$ regularization penalizes bias vectors, which do not contribute to the high-frequency noise in the derivative.
Numerical experiments in \autoref{sec:experiments} indicate that SNFW-ROM trained with a large value of $\gamma$ consistently outperforms the other benchmark models.
\noteB{
A comparison of regularization approaches is presented in \autoref{sec:appendix-regularization}.
}

%%%%%%%%%%%%%%%%%%%%%%%%%%%%%%%%%%%%%%%%%%%%%%%%%%%%%%%%%%%%%%%%%%%
\section{Evaluation of reduced state dynamics}
\label{sec:dynamics}
%%%%%%%%%%%%%%%%%%%%%%%%%%%%%%%%%%%%%%%%%%%%%%%%%%%%%%%%%%%%%%%%%%%

\begin{figure}
    \centering

    \subfigure[\label{fig:commutative}]{
    \includegraphics[height=0.3\textwidth,clip]{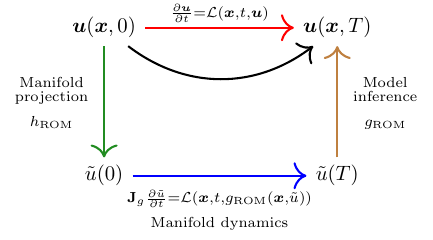}
    }
    \subfigure[\label{fig:manifold_dynamics}]{
    \includegraphics[height=0.3\textwidth,clip]{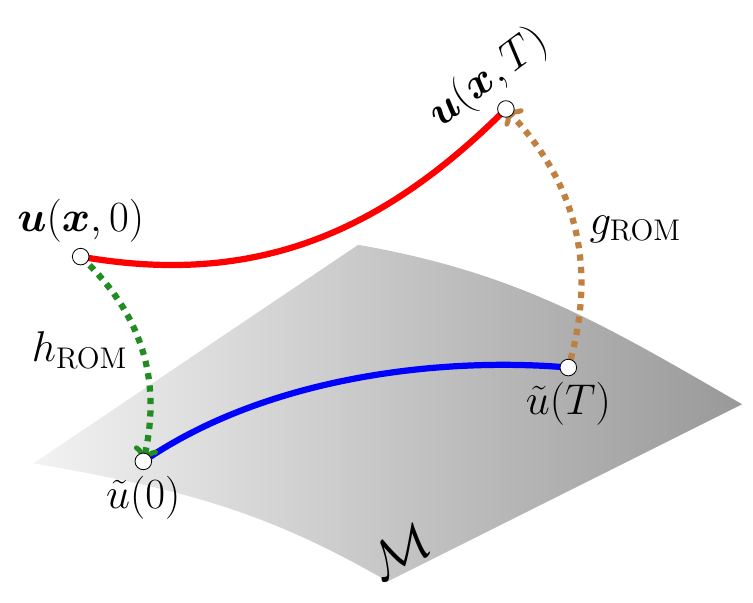}
    }
    \vspace{-1em}
    \caption{
    The online dynamics evaluation is described (a) with a commutative diagram, and (b) with a schematic diagram.
    Reduced dynamics are evaluated in three steps.
    (1) Manifold projection: The PDE initial condition $\vect{u}(0)$ is projected onto the ROM manifold $\mathcal{M}$ to obtain the corresponding ROM state vector $\tilde{u}(0)$.
    (2) Manifold dynamics: $\tilde{u}$ is time-marched on $\mathcal{M}$ till $t=T$ following the dynamics of the governing PDE to obtain $\tilde{u}(T)$.
    (3) Model inference: $\tilde{u}(T)$ is inferred for analysis and post-processing as $\vect{u}(\vect{x}, T) = g_\theta(\tilde{u}(T))$.
    }
    \label{fig:online}
\end{figure}

This section describes the online stage of reduced state dynamics with Galerkin projection.
As illustrated in \autoref{fig:online},
the dynamics evaluation of the reduced system is done in three steps:
(1) \text{Manifold projection:} we project the initial condition $\vect{u}(\vect{x}, 0)$ to \autoref{eqn:prob} onto the ROM manifold and obtain the corresponding ROM state vector $\tilde{u}(0)$.
This step is performed only once.
(2) \text{Manifold dynamics:} $\tilde{u}(t)$ is evolved from $t=0$ to $t=T$ following the Galerkin projection approach \cite{carlberg_galerkin_2016}.
%This involves evaluating the model, its derivatives, and Jacobian at a small number of points, $X_\text{proj} \subset \Omega$.
(3) \text{Model inference:} once time-evolution is complete, the PDE solution at time $t=T$ is evaluated as $g_\text{ROM}(\vect{x}, \tilde{u}(T))$ at query points $\vect{x}\in\Omega$.
In the rest of this section, we omit dependence on the parameter 
vector $\vect{\mu}$ for ease of exposition.

%====================================================%
\subsection{Manifold projection}
\label{subsec:manifold-projection}
%====================================================%

As illustrated in \autoref{fig:online}, the initial condition
$\vect{u}(0) = \vect{u}_0(\vect{x})$ to \autoref{eqn:prob}
is projected onto the manifold with $h_\text{ROM}$ to obtain the corresponding ROM state vector $\tilde{u}(0)$.
This step is only done once to initialize the dynamics evaluation.
% POD-ROM
For POD-ROMs, the projection function is a matrix-vector product as described in \autoref{eqn:POD-h}.
% CAE-ROM
For CAE-ROM, a preliminary approximation to $\tilde{u}(0)$ is given by the encoder prediction $e_{\theta_e}(\dvect{u}(0))$, where $\dvect{u}$ is the FOM state vector corresponding to the initial condition.
Our experiments indicate that $e_{\theta_e}(\dvect{u}(0))$ may not be the optimal projection.
As such, the accuracy of the dynamics evaluation can be significantly improved by employing Gauss-Newton iteration to further optimize the return value.
CAE-ROM thus solves the nonlinear system in \autoref{eqn:CAE-proj} with $N_\text{FOM}$ equations for $N_\text{ROM}$ unknowns with Gauss-Newton iteration where the initial guess is taken to be the encoder prediction $e_{\theta_e}(\dvect{u}(0))$.
Further details are presented in \aref{subsec:appendix-cae-rom-projection}.

% SNF-ROM
For SNF-ROM, a preliminary approximation to $\tilde{u}(0)$ is given by the intrinsic ROM manifold $\Xi_\varrho(0)$.
Our experiments indicate that using Gauss-Newton iteration to solve \autoref{eqn:neuralfield-proj} after the process leads to marginal improvements in the accuracy of the dynamics evaluation.
As such, we select a small set of points $X_\text{proj} \subset \Omega$ where \autoref{eqn:neuralfield-proj-approx} is satisfied.
The choice of points in $X_\text{proj}$ is not restricted to $X_\text{FOM}$ and can be sampled anywhere in $\Omega$.
Although only $N_\text{ROM}$ points are needed for solving \autoref{eqn:neuralfield-proj-approx-xproj},
the system is typically over-determined \cite{chen_crom_2023} with $N_\text{ROM} < \abs{X_\text{proj}} << N_\text{FOM}$ where $\abs{X_\text{proj}}$ is the number of points in $X_\text{proj}$.
The resulting system is much smaller (see \autoref{eqn:neuralfield-proj}), with $\abs{X_\text{proj}}$ equations and $N_\text{ROM}$ unknowns, and solved with Gauss-Newton iteration where the initial guess is given by $\Xi_\varrho$.

%====================================================%
\subsection{Manifold dynamics}
\label{subsec:latent-dynamics}
%====================================================%

Once the initial ROM state vector ${\tilde{u}}(0)$ is found, we begin time-evolution on the reduced manifold as illustrated in \autoref{fig:manifold_dynamics}.
For time-marching of ROM state vector, we substitute the ROM ansatz,
$\vect{u}_\text{ROM}(\vect{x}, t) = g_\text{ROM}(\vect{x}, {\tilde{u}}(t))$,
in \autoref{eqn:prob} to get
\eqn{
    \label{eqn:dynamics-g-ode}
    \ppp{t} g_\text{ROM} (\vect{x}, {\tilde{u}}(t) )
    &=
    \mathcal{L}\left(
    \vect{x}, t, g_\text{ROM}(\vect{x}, {\tilde{u}}(t)) \right).
}
As the ROM state at time $t$ is known,
\eqn{
    \vect{f}(\vect{x}, t, {\tilde{u}}(t))
    \defeq
    \mathcal{L}\left(
    \vect{x}, t, g_\text{ROM}(\vect{x}, {\tilde{u}}(t))
    \right)
}
can be evaluated anywhere in $\Omega$.
%===============%
% RHS - derv
%===============%
Evaluating $\mathcal{L}$ involves computing spatial derivatives of $\vect{u}$. As mentioned earlier, CAE-ROM computes spatial derivatives at $X_\text{FOM}$ by applying a low-order finite difference stencil \cite{chen_crom_2023}.
%Neural field ROMs compute spatial derivatives with a carefully chosen finite difference stencil to dampen noise in the derivative field \cite{chen_crom_2023}.
As SNF-ROM is inherently smooth, we can compute accurate spatial derivatives with forward mode AD as verified using numerical experiments.

We aim to satisfy \autoref{eqn:dynamics-g-ode} at points in $X_\text{proj} \subset \Omega$.
As such, we refer to $X_\text{proj}$ as the set of hyper-reduction collocation points.
For CAE-ROM, $X_\text{proj} = X_\text{FOM}$ as CAE architectures do not implement hyper-reduction, and predict the entire solution field at inference time \cite{lee_model_2020}.
As such, the cost for CAE-ROM scales with that of the FOM.
In comparison, $X_\text{proj}$ can be a small subset of $\Omega$ for SNF-ROM, which leads to substantial computational savings.
We restrict the semi-discretized system (\autoref{eqn:dynamics-g-ode}) to $X_\text{proj}$ and write it in matrix-vector notation as
\eqn{
    \label{eqn:dynamics-J-ode-matvec}
    \mathbf{J}_{g}({\tilde{u}}(t)) \cdot
    \ddd{t} {\tilde{u}}(t)
    &= \dvect{f}(t, {\tilde{u}}(t)),
}
where
\eqn{
    \label{eqn:g_jacobian}
    \mathbf{J}_{g}({\tilde{u}}; X_\text{proj})
    = \mat{
        \vdots\\
        \ppp{{\tilde{u}}} g_{\text{ROM}}(\vect{x}, {\tilde{u}})\\
        \vdots
    }_{\vect{x} \in X_\text{proj}}
}
is the
Jacobian matrix
(with $m\times\abs{X_\text{proj}}$ rows and $N_\text{FOM}$ columns)
of $g_\text{ROM}$ with respect to $\tilde{u}$, and
\eqn{
    \dvect{f}(t, {\tilde{u}}; X_\text{proj}) = \mat{
        \vdots\\
        \vect{f}(\vect{x}, t, {\tilde{u}})\\
        \vdots
    }_{\vect{x} \in X_\text{proj}}.
}
%===============%
We follow the time-continuous residual minimization methodology \cite{carlberg_galerkin_2016,lee_model_2020, kim_fast_2022} that leads to Galerkin-optimal ROM solutions.
The Galerkin projection scheme seeks to project the time-continuous ODE system \autoref{eqn:dynamics-J-ode-matvec} onto the reduced manifold $\mathcal{M}$.
%That is, we seek $\tilde{f}(t, \tilde{u})\in\R^{N_\text{ROM}}$ that describes the dynamics of the ROM state as
%\eqn{
%    \label{eqn:dpdt}
%    \ddd{t}{\tilde{u}}(t) = \tilde{f}(t, \tilde{u}).
%}
To derive the Galerkin projection scheme, we close the overdetermined system in \autoref{eqn:dynamics-J-ode-matvec} by minimizing the squared norm of its residual.
That is, we seek velocity $\dfrac{\d}{\d t}{\tilde{u}(t)}$ such that
\eqn{
    \label{eqn:dynamics-vel-argmin}
    \ddd{t} {\tilde{u}}(t) = 
    \argmin_{\dvect{v}\in\R^{N_\text{ROM}}}
    \norm{
        \mathbf{J}_{g}({\tilde{u}}(t)) \cdot \dvect{v}
        -
        \dvect{f}(t, {\tilde{u}}(t))
    }_2^2.
}
The linear least-squares problem in \autoref{eqn:dynamics-vel-argmin} can be solved exactly to obtain the system
\eqn{
    \label{eqn:dynamics:reduced-system}
    \ddd{t}{\tilde{u}}(t) =
    \mathbf{J}_{g}({\tilde{u}}(t))^\dag
    \cdot
    \dvect{f}(t, {\tilde{u}}(t))
}
where $\mathbf{J}^\dag = (\mathbf{J}^T \cdot \mathbf{J})^{-1}\mathbf{J}^T$ is the pseudo-inverse of matrix $\mathbf{J}$.
%This is equivalent to projecting the residual of \autoref{eqn:dynamics-J-ode-matvec} onto the tangent plane of the manifold $\mathcal{M}$ at point ${\tilde{u}}(t)$:
%\eqn{
%    \left\langle
%        \mathbf{J}_{g}({\tilde{u}}(t)),\,
%        \mathbf{J}_{g}({\tilde{u}}(t)) \cdot \ddd{t}{\tilde{u}}(t) -
%        \dvect{f}(t, {\tilde{u}}(t))
%    \right\rangle
%    = 0
%}
In practice, the pseudo-inverse is not computed,
but the RHS of \autoref{eqn:dynamics:reduced-system} is computed with QR factorization.
%but the size $N_\text{ROM}\times N_\text{ROM}$ linear system
%\eqn{
%    \left(
%    \mathbf{J}_{g}({\tilde{u}}(t))^T
%    \mathbf{J}_{g}({\tilde{u}}(t))
%    \right)
%    \cdot
%    \tilde{f}(t, {\tilde{u}}(t))
%    =
%    \mathbf{J}_{g}({\tilde{u}}(t))^T
%    \dvect{f}(t, {\tilde{u}}(t))
%}
%is solved with QR factorization for $\tilde{f}(t, \tilde{u}(t))$.
As $N_\text{ROM}$ is small, this computation introduces minimal overhead when the projection set $X_\text{proj}$ is small \cite{chen_crom_2023}.
The reduced system in \autoref{eqn:dynamics:reduced-system} is then marched in time with an explicit time integrator such as the Euler or Runge-Kutta (RK) method.

% LSPG
\noteB{
An alternative time-evolution strategy, called least-squares Petrov-Galerkin (LSPG) \cite{carlberg_galerkin_2016}, is obtained by applying a time-discretization to \autoref{eqn:dynamics-J-ode-matvec}.
The problem is then evolved by solving the nonlinear system for $\tilde{u}$ at the next time-step with nonlinear least squares.
We consider explicit time-integrators in this paper, for which LSPG has been shown to be equivalent to Galerkin projection \cite{carlberg_galerkin_2016}.
%Our internal experiments reveal no difference in the prediction between LSPG and Galerkin Projection.
%However, LSPG requires much greater wall-time due to the need to solve a nonlinear system at each time-step.
As such, we focus only on Galerkin projection method for the rest of this paper.
}

%====================================================%
\subsection{Model inference}
\label{subsec:latent-final}
%====================================================%
ROM state vectors can be saved at different times throughout the time-marching while
incurring significantly lower memory cost than FOMs as $N_\text{ROM}$ is typically much smaller than $N_\text{FOM}$.
Saved ROM snapshots $\tilde{u}(t; \vect{\mu})$ can then be evaluated with $g_\text{ROM}$ to produce corresponding PDE solution fields at query points $\vect{x} \in \Omega$.
\autoref{fig:manifold_dynamics} illustrates
the PDE solution at the final time-step being obtained as
\eqn{
    \vect{u}(\vect{x}, T) = g_\text{ROM}(\vect{x}, \tilde{u}(T)).
}
This evaluation is typically done in a post-processing step after the solution is computed.

%%%%%%%%%%%%%%%%%%%%%%%%%%%%%%%%%%%%%%%%%%%%%%%%%%%%%%%%%%%%%%%%%%%
\section{Numerical experiments}
\label{sec:experiments}
%%%%%%%%%%%%%%%%%%%%%%%%%%%%%%%%%%%%%%%%%%%%%%%%%%%%%%%%%%%%%%%%%%%

In this section, we validate the SNFL-ROM and SNFW-ROM on several advection-dominated PDE problems and compare them to POD-ROMs and CAE-ROMs \cite{lee_model_2020}.
% CROM dynamics evaluation is sensitive
%\st{
%Another new approach that relies on neural fields, CROM \cite{chen_crom_2023}, was also tested.
%However, we found the dynamics evaluation in CROM to be sensitive to the approach used for numerical derivative computation while also diverging for reduced state vector trajectories within the training dataset. 
%}
% CoLoRa comparison
The concurrent work in \cite{berman_colora_2024} employs a continuous low-rank architecture to develop an alternate grid-free neural representation for reduced order modeling.
Future work may consider a detailed comparison of SNF-ROM with this approach.

% data generation
We use a Fourier spectral PDE solver \texttt{FourierSpaces.jl} \cite{puri_2022_fourierspacesjl} to generate the FOM datasets.
A strong stability-preserving Runge-Kutta time integrator, available in package \texttt{OrdinaryDiffEq.jl} \cite{rackauckas_differentialequationsjl_2017}, is used for temporal discretization.
% training
For each experiment, $X_\text{FOM}$ is a uniform rectilinear grid in the problem domain, and $T_\text{FOM}$ are uniformly sampled from $t=0$ through $t=T$.
Details of the dataset for each test case are presented in \autoref{tab:dataset}.
Note that problems with an intrinsic manifold dimension of $1$ are reproductive examples where the parameter $\vect{\mu}$ is kept constant, implying that the ROM dynamics are evaluated on the same trajectory as the model is trained.
Such examples help to disambiguate errors from parameter interpolation and errors from deviations in the dynamics evaluation.

\begin{table}
    \centering
    \caption{
        Description of the PDE problem and FOM dataset for each test case.
    }
    \vspace{-1em}
    \begin{tabular}{|c|c|c|c|c|c|}
        \hline
        Problem &
        \shortstack{Intrinsic manifold\\ dimension} &
        Domain ($\Omega$) &
        Grid size ($N_\text{FOM}$) &
        End time ($T$) &
        \shortstack{Number of\\ snapshots ($N_t$)} \\
        \hline
        Advection 1D & $1$ & $[-1, 1)$         & $128$          & $4  $ & $500$ \\\hline
        Advection 2D & $1$ & $[-1, 1)^2$       & $128\times128$ & $4  $ & $500$ \\\hline
        Burgers 1D   & $2$ & $[0, 2)$          & $1,024$        & $0.5$ & $500$ \\\hline
        Burgers 2D   & $2$ & $[-0.25, 0.75)^2$ & $512\times512$ & $0.5$ & $500$ \\\hline
        KS 1D        & $1$ & $[-\pi, \pi)$     & $256$          & $0.1$ & $1,000$ \\\hline
    \end{tabular}
    \label{tab:dataset}
\end{table}

% time integrator
For the dynamics evaluation for ROMs, we use a first order explicit Euler time integrator with a uniform step size of
\eqn{
    \label{eqn:dt-baseline}
    \Delta t_{0} = \frac{T}{N_t - 1}.
}
Our numerical experiments revealed no substantial improvement in performance with higher-order time-integration schemes such as RK4.
The accuracy of all ROMs deteriorates with the time-step size.
Our experiments in this section reveal that the accuracy of SNF-ROM does not significantly degrade with increasing $\Delta t$.
Taking large time-steps can thus lead to a substantial speedup without compromising accuracy of the method.

% X_proj
Unless otherwise noted, we set $X_\text{proj}$ to be the same as $X_\text{FOM}$ in order to carry out a fair comparison with the CAE-ROM, which does not include hyper-reduction \cite{kim_fast_2022} in its current implementation.
To reduce the cost of the FOM computation, we implement hyper-reduction by choosing the set of collocation points $X_\text{proj}$ such that $\abs{X_\text{proj}} < \abs{X_\text{FOM}}$.
As such, speedup in the online evaluation is reported for \autoref{subsec:exp2} and \autoref{subsec:exp4}.
In these experiments, $X_\text{proj}$ is obtained by uniformly subsampling the FOM grid $X_\text{FOM}$.
Future work may consider incorporating sophisticated strategies for obtaining $X_\text{proj}$ such as the greedy approach developed in \cite{chen_crom_2023}.

% spatial derivatives
We efficiently compute Jacobian matrices by propagating dual numbers through $g_\text{ROM}$ in the forward mode AD procedure.
As such, the cost of computing $\mathbf{J}_g(\tilde{u}; X_\text{proj})$ is only $N_\text{ROM}$ times that of a single forward pass, $g_\text{ROM}$, evaluated on $X_\text{proj}$. This cost is independent of the FOM computation when hyper-reduction is employed.
Spatial derivatives for POD-ROM and CAE-ROM are computed with a second order central finite difference stencil on the $X_\text{FOM}$ grid.
For SNFL-ROM and SNFW-ROM, spatial derivatives are computed grid-free with forward mode AD.

% hyperparameter
The choice of hyperparameters for all models is presented in \autoref{tab:hyperparameters} for each test case.
A common drawback of machine learning methods is that significant time and computing resources are spent optimizing hyperparameters to produce good results for each problem.
We find that the choice of hyperparameters for SNF-ROM remains consistent across a broad range of problems, which is an additional advantage.

\begin{table}
    \centering
    \caption{
        Hyperparameters of POD-ROM, CAE-ROM, SNFL-ROM, and SNFW-ROM for each test case.
    }
    \vspace{-1em}
    \begin{tabular}{| *{9}{c|} }
    
    \hline

    &
    POD &
    \multicolumn{3}{c|}{CAE} &
    \multicolumn{4}{c|}{SNF} \\

    \cline{2-9}
    %\hline

    {Problem} &
    $N_\text{ROM}$ &
    $N_\text{ROM}$ &
    $N_\text{param}$ &
    $\sigma$ &
    $N_\text{ROM}$ &
    $N_\text{param}$ &
    $\alpha$ (SNFL) &
    $\gamma$ (SNFW) \\

    \hline

    Advection 1D &
    $8$ &                    % POD
    $2$ &                    % CAE - N_ROM
    $21\mbox{k}$ ($w=32$) &  % CAE - N_DEC
    $\mbox{elu}$ &           % CAE - sigma
    $2$ &                    % SNF - N_ROM
    $21\mbox{k}$ ($w=64$) &  % SNF - N_DEC
    $10^{-4}$ &              % SNF - alpha
    $10^{-2}$                % SNF - gamma
    \\\hline

    Advection 2D &
    $8$ &                    % POD
    $2$ &                    % CAE - N_ROM
    $594\mbox{k}$ ($w=64$) & % CAE - N_DEC
    $\tanh$ &                % CAE - sigma
    $2$ &                    % SNF - N_ROM
    $82\mbox{k}$ ($w=128$) & % SNF - N_DEC
    $10^{-4}$ &              % SNF - alpha
    $10^{-2}$                % SNF - gamma
    \\\hline

    Burgers 1D &
    $8$ &                    % POD
    $2^*$ &                  % CAE - N_ROM
    $115\mbox{k}$ ($w=64$) & % CAE - N_DEC
    $\tanh$ &                % CAE - sigma
    $2^*$ &                  % SNF - N_ROM
    $82\mbox{k}$ ($w=128$) & % SNF - N_DEC
    $10^{-4}$ &              % SNF - alpha
    $10^{-2}$                % SNF - gamma
    \\\hline
    
    Burgers 2D &
    $16$ &                   % POD
    $2^*$ &                  % CAE - N_ROM
    $800\mbox{k}$ ($w=64$) & % CAE - N_DEC
    $\mbox{elu}$ &           % CAE - sigma
    $2^*$ &                  % SNF - N_ROM
    $82\mbox{k}$ ($w=128$) & % SNF - N_DEC
    $10^{-4}$ &              % SNF - alpha
    $10^{-2}$                % SNF - gamma
    \\\hline
    
    KS 1D &
    $2$ &                    % POD
    $1^*$ &                  % CAE - N_ROM
    $82\mbox{k}$ ($w=64$) & % CAE - N_DEC
    $\mbox{elu}$ &           % CAE - sigma
    $1^*$ &                  % SNF - N_ROM
    $82\mbox{k}$ ($w=128$) & % SNF - N_DEC
    $10^{-7}$ &              % SNF - alpha
    $10^{-2}$                % SNF - gamma
    \\\hline
    
    \end{tabular}
    \begin{tablenotes}
        \item
        Note: ``$*$'' indicates that the ROM dimension is equal to the intrinsic manifold dimension of the test problem;
        $N_\text{param}$ is the number of parameters in $g_\text{ROM}$;
        $w$ is the width of the hidden layer in SNF-ROM and the depth of the convolutional kernels for CAE-ROM; and
        $\sigma$ is the activation function.
    \end{tablenotes}
    \label{tab:hyperparameters}
\end{table}

% Error metric
\noteB{
We quantify the projection error for each ROM method as
\eqn{
    \label{eqn:error_proj}
    e_\text{proj} &=
    \left(
        \frac{1}{
            \abs{T_\text{FOM}}
            \abs{M_\text{ROM}}
        }
        \sum_{\substack{
            t \in T_\text{FOM} \\
            \vect{\mu}\in M_\text{FOM}}
        }
        \frac{
            \norm{
                \dvect{u}(t; \vect{\mu})
                -
                g_\text{ROM} \circ h_\text{ROM}(\dvect{u}(t; \vect{\mu}))
            }_2^2
        }{
            \norm{
                \dvect{u}(t; \vect{\mu})
            }_2^2
        }
    \right)^{1/2}
    .
}
This metric assesses the ability of a model order reduction approach to capture the solution manifold for a PDE problem.
}
The relative error as a function of $(\vect{x}, t; \vect{\mu})$ in ROM predictions is quantified as
\eqn{
    \label{eqn:error_xt}
    \epsilon(\vect{x}, t; \vect{\mu}) &=
    \frac{
        \norm{
            \vect{u}_\text{FOM}(\vect{x}, t; \vect{\mu})
            -
            \vect{u}_\text{ROM}(\vect{x}, t; \vect{\mu})
        }_2
    }{
        \sqrt{
            \frac{1}{\abs{X_\text{FOM}}}
            \sum_{\vect{x}\in X_\text{FOM}}
            \norm{
                \vect{u}_\text{FOM}(\vect{x}, t; \vect{\mu})
            }_2^2
        }
    }
    ,
}
where $\vect{u}_{\text{FOM}}$ and $\vect{u}_{\text{ROM}}$ are spatiotemporal snapshots of the FOM and ROM solutions, respectively.
The evolution of relative error in ROM predictions is quantified using
\eqn{
    \label{eqn:error_t}
    \varepsilon(t; \vect{\mu}) &=
    \sqrt{
        \frac{1}{\abs{X_\text{FOM}}}
        \sum_{\vect{x}\in X_\text{FOM}}
        \epsilon(\vect{x}, t; \vect{\mu})^2
    }
    =\left(
        \sum_{\vect{x}\in X_\text{FOM}}
        \norm{
            \vect{u}_\text{FOM}(\vect{x}, t; \vect{\mu})
            -
            \vect{u}_\text{ROM}(\vect{x}, t; \vect{\mu})
        }_2^2
    \right)^{1/2}
    \Biggr/
    \left(
        \sum_{\vect{x}\in X_\text{FOM}}
        \norm{
            \vect{u}_\text{FOM}(\vect{x}, t; \vect{\mu})
        }_2^2
    \right)^{1/2}
    .
}
%Finally, the relative error in a ROM evaluation is given as
%\eqn{
%    \label{eqn:error_}
%    e(\vect{\mu}) &=
%    \sqrt{
%        \frac{1}{\abs{T_\text{FOM}}}
%        \sum_{\vect{t}\in T_\text{FOM}}
%        \varepsilon(t; \vect{\mu})^2
%    }
%    =\left(
%        \sum_{\substack{\vect{x}\in X_\text{FOM} \\ t\in T_\text{FOM}}}
%        \norm{
%            \vect{u}_\text{FOM}(\vect{x}, t; \vect{\mu})
%            -
%            \vect{u}_\text{ROM}(\vect{x}, t; \vect{\mu})
%        }_2^2
%    \right)^{1/2}
%    \Biggr/
%    \left(
%        \sum_{\substack{\vect{x}\in X_\text{FOM} \\ t\in T_\text{FOM}}}
%        \norm{
%            \vect{u}_\text{FOM}(\vect{x}, t; \vect{\mu})
%        }_2^2
%    \right)^{1/2}
%    .
%}
Details of the training procedure for SNF-ROM are presented in \aref{sec:appendix-snf-rom-train}.
% implementation
The training and inference pipelines for all models are implemented in the Julia programming language \cite{bezanson_julia_2012}
using the Lux \cite{pal_efficient_2023} deep learning framework using the \texttt{CUDA.jl} library \cite{besard_effective_2019} for GPU acceleration.
The \texttt{ZygoteAD} \cite{Zygote.jl-2018} is used for backpropagation during training,
and forward mode AD is implemented with the package \texttt{ForwardDiff.jl} \cite{revels_forward-mode_2016}.
% hardware
All FOM and ROM calculations have been carried out on a single Nvidia $2080$ Ti GPU with $11\mbox{GiB}$ of VRAM.

%====================================================%
\subsection{1D scalar advection test case}
\label{subsec:exp1}
%====================================================%

\begin{figure}[t]
    \centering
    \includegraphics[width=\textwidth]{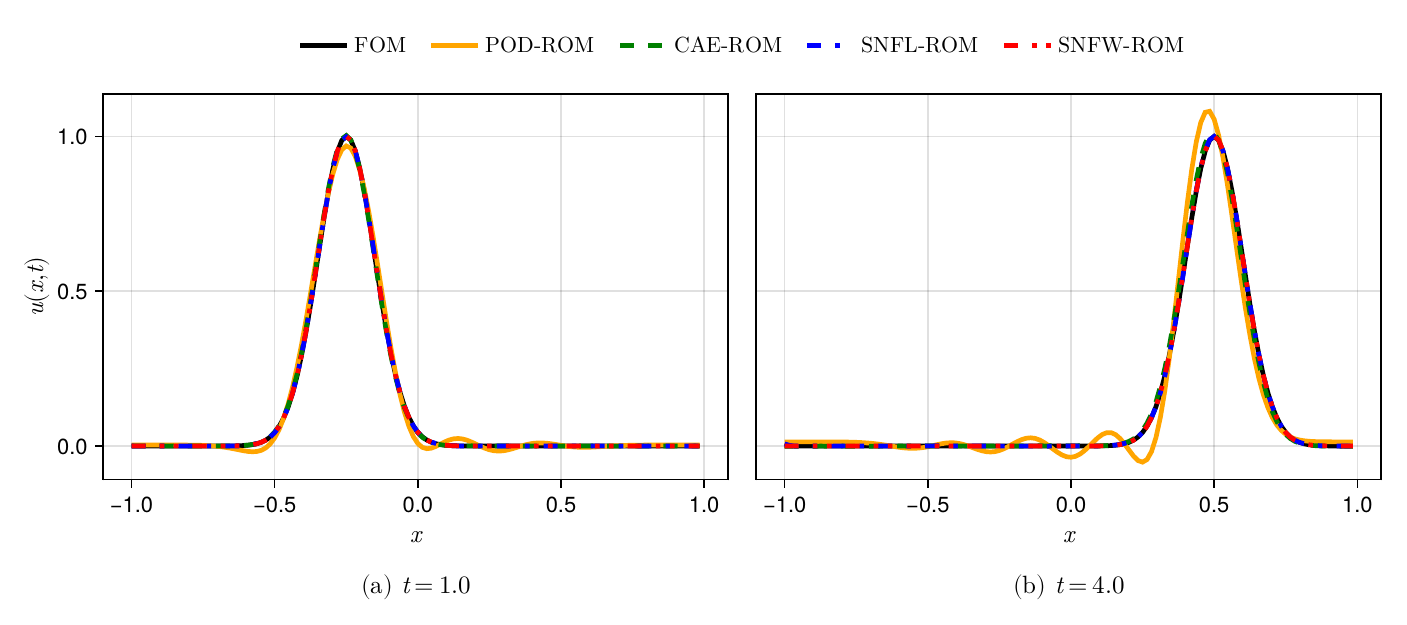}
    \vspace{-2em}
    \caption{
    1D scalar advection test case:
    comparison of different ROM predictions with the FOM solution.
    The predicted solutions for CAE-ROM, SNFW-ROM, and SNFL-ROM overlap and match the FOM solution, whereas those for POD-ROM suffer from large oscillations.
    }
    \label{fig:exp1f1}
\end{figure}

\begin{figure}[t]
    \centering
    \includegraphics[width=0.6\textwidth]{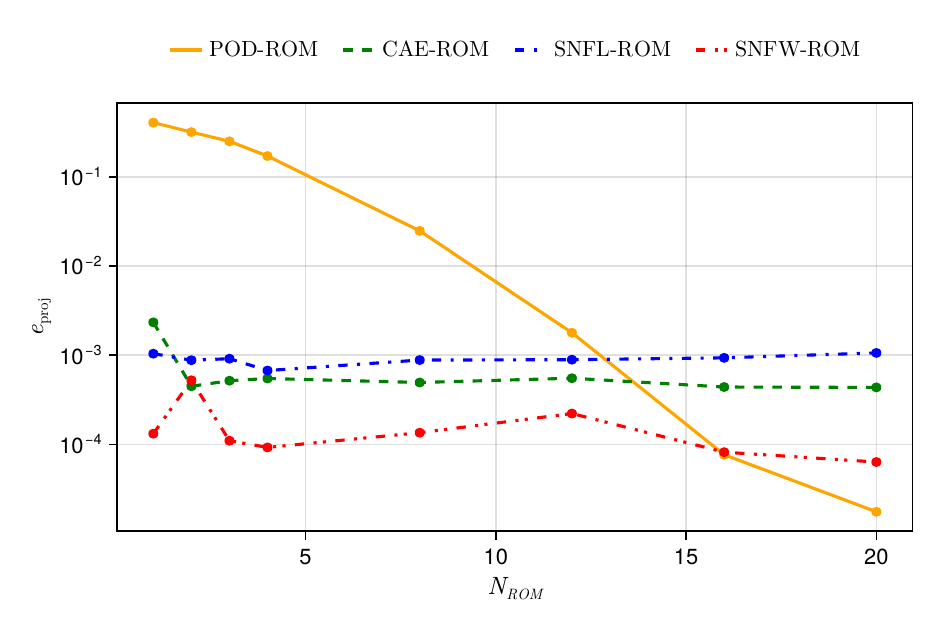}
    \caption{
        \noteB{
        1D scalar advection test case: comparison of projection error $e_\text{proj}$ of different ROMs as a function of ROM dimension $N_\text{ROM}$.
        The three nonlinear ROMs adequately capture the solution manifold with $N_\text{ROM} = 1$ which is equal to the intrinsic manifold dimension of the problem.
        Comparatively, the linear POD-ROM requires $N_\text{ROM} = 16$ modes to capture the solution manifold for this case.
        }
    }
    \label{fig:exp_nrom}
\end{figure}

\begin{figure}[!ht]
    \centering
    \includegraphics[width=\textwidth]{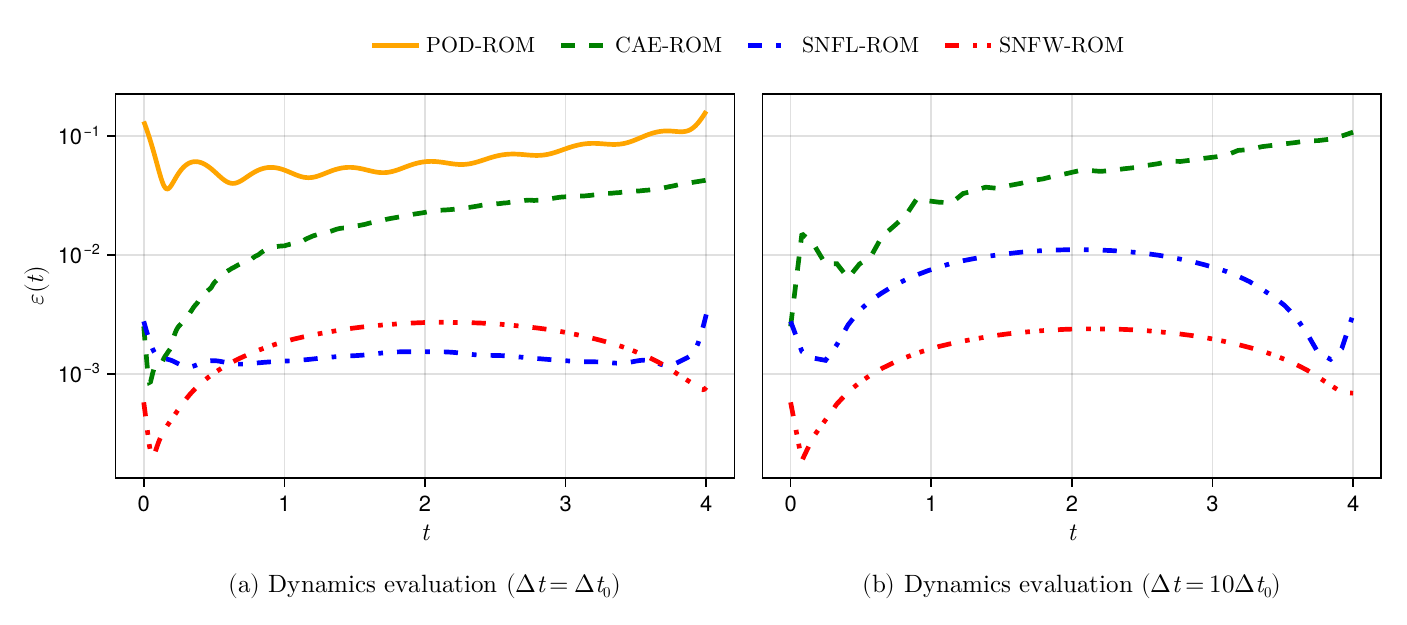}
    \vspace{-2em}
    \caption{
        1D scalar advection test case:
        error evolution of ROM predictions in time for dynamics evaluation with (a) $\Delta t = \Delta t_0$, and (b) $\Delta t = 10\Delta t_0$.
        The error for CAE-ROM grows with time for both cases, while the error for SNFW-ROM remains consistently low.
        The curve for POD-ROM is omitted from (b) due to large error values.
    }
    \label{fig:exp1f2}
\end{figure}

\begin{figure}[!ht]
    \centering
    \includegraphics[width=1.0\linewidth]{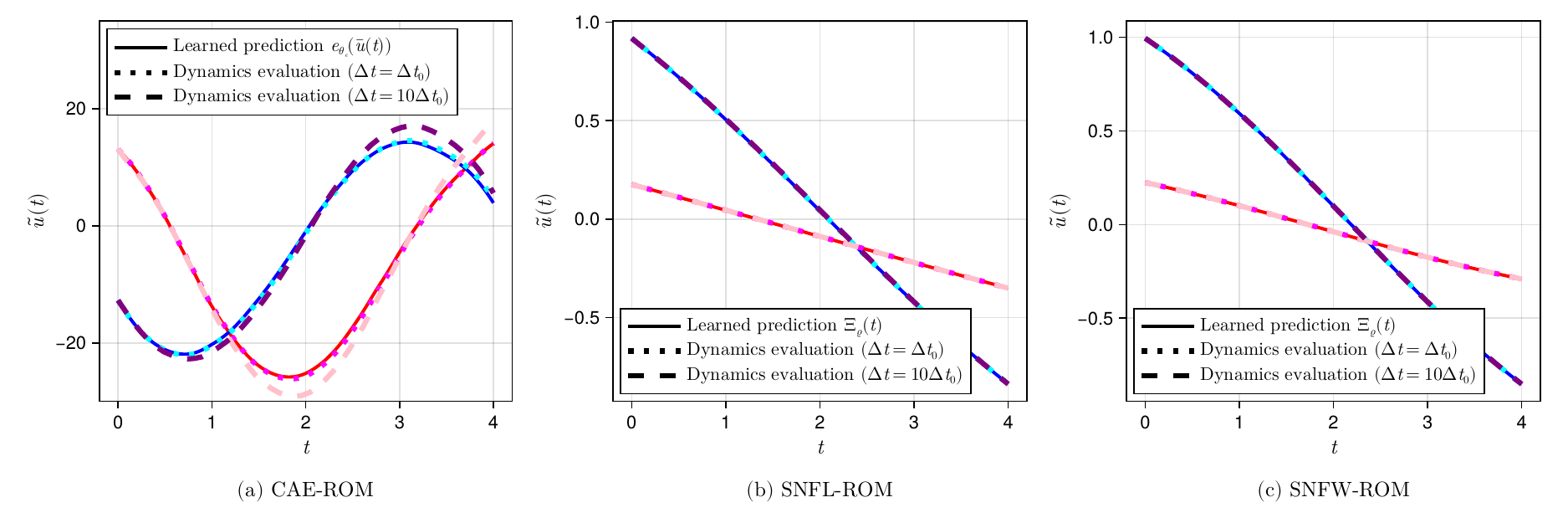}
    \vspace{-2em}
    \caption{
    1D scalar advection test case:
    time-evolution of the ROM state vector
    $\tilde{u}(t) = \mat{\tilde{u}_1(t) &\hspace{-0.5em} \tilde{u}_2(t)}^T$.
    The blue and red curves correspond to the first ($\tilde{u}_1$) and second $(\tilde{u}_2)$ components respectively.
    For CAE-ROM, $\tilde{u}(t)$ from the dynamics evaluation deviates from the learned trajectory, whereas
    the curves from dynamics evaluations and the learned trajectory overlap for both SNFL-ROM and SNFW-ROM.
    }
    \label{fig:exp1f5}
\end{figure}

The first test case involves solving the 1D scalar advection equation,
\eqn{
    \label{eqn:advection1D}
    \frac{\p u}{\p t} + c\frac{\p u}{\p x} = 0,
}
which governs the transport of a scalar $u$.
The advective velocity is selected to be $c = 0.25$. A Gaussian signal, 
\eqn{
    u_0(x) = \exp\left\{
        \frac{-(x - \mu_0)^2}{2\sigma_0^2}
        \right\}
    ,
}
with $\mu_0 = -0.5$ and $\sigma_0 = 0.1$, is used to initialize the problem.
As advection is the only transport mechanism for the scalar in this test case, it suffers from a slow decay of the Kolmogorov $n$-width, thereby motivating the need for nonlinear model reduction approaches.

A comparison of ROM predictions with the FOM solution is shown in \autoref{fig:exp1f1} for $t=1$ and $t=4$.
The figure indicates that the nonlinear ROMs give close predictions to FOM in both instances.
On the other hand, POD-ROM leads to oscillations in the solution that become more prominent near the wave.

\noteB{
To assess the model order reduction capability of different ROMs, we compute the projection error as a function of ROM dimension $N_\text{ROM}$ in \autoref{fig:exp_nrom}.
As expected, the linear POD-ROM requires a large number of modes to capture the solution manifold.
This is because advection dominated problems exhibit a slow decay in their Kolmogorov $n$-width.
In contrast, the three nonlinear ROMs are able to adequately capture the solution manifold with $N_\text{ROM} = 1$, which is the intrinsic manifold dimension of this problem.
Specifically, SNFW-ROM best captures the solution manifold for small $N_\text{ROM}$.
We also note that increasing $N_\text{ROM}$ beyond the intrinsic manifold dimension of the problem does not lead to a significant improvement in the performance of nonlinear models.
}

A more detailed analysis of the differences between the ROM approaches requires assessing the temporal variation of the error, as defined in \autoref{eqn:error_t}.
\autoref{fig:exp1f2} presents the error evolution of ROMs from a dynamics evaluation with $\Delta t = \Delta t_0$, and a dynamics evaluation with a $\Delta t = 10\Delta t_0$.
POD-ROM exhibits a large error at all times, indicating that a larger ROM representation is required to capture physics accurately.
The curve for POD-ROM is omitted from \autoref{fig:exp1f2}(b) due to large errors.
In contrast, the nonlinear ROMs yield lower errors with a much smaller $N_\text{ROM}$.
Despite a low initial value, the error in CAE-ROM grows steadily with time for both $\Delta t$, and reaches close to the POD-ROM error levels.
The two SNF-ROM variants produce much lower errors for both $\Delta t$, though we observe a dramatic rise in error for SNFL-ROM with the larger time step size between $t=0.5$ and $t=3.5$. In contrast,
SNFW-ROM produces consistently low errors for both $\Delta t$.

We analyze the trajectories of the ROM state vector for the nonlinear ROMs in \autoref{fig:exp1f5} to assess deviation between predictions of $\tilde{u}(t)$ learned during training (solid lines) and
$\tilde{u}(t)$ obtained from dynamics evaluations (dotted lines correspond to a dynamics evaluation with $\Delta t = \Delta t_0$, and dashed lines correspond to a dynamics evaluation with $\Delta t = 10\Delta t$).
The first and second components of
$\tilde{u}(t) = \mat{\tilde{u}_1(t) &\hspace{-0.5em} \tilde{u}_2(t)}^T$
are represented by blue and red curves respectively.
\autoref{fig:exp1f5} indicates that the three nonlinear ROMs learn ROM state trajectories that are smooth and continuous.
SNFL-ROM and SNFW-ROM explicitly force this behavior by modeling $\tilde{u}$ as a simple and smooth function of $t$, which follows the expected trajectory accurately.
For CAE-ROM, despite the initial proximity to the learned trajectory, ROM states from the dynamics evaluation with $\Delta t = \Delta t_0$ deviate from the learned trajectory.
This deviation corresponds to the growth in error with time for CAE-ROM presented in \autoref{fig:exp1f2} although further studies need to be conducted to ascertain causality.
Furthermore, the ROM state trajectories for CAE-ROM obtained using a $10\times$ larger $\Delta t$ quickly deviate from the learned trajectory.
This experiment illustrates that the dynamics evaluations of SNFL-ROM and SNFW-ROM are more accurate than those of other ROMs and support taking larger time steps.

%====================================================%
\subsection{2D scalar advection test case}
\label{subsec:exp2}
%====================================================%

\begin{figure}[!ht]
    \centering
    \includegraphics[width=\textwidth]{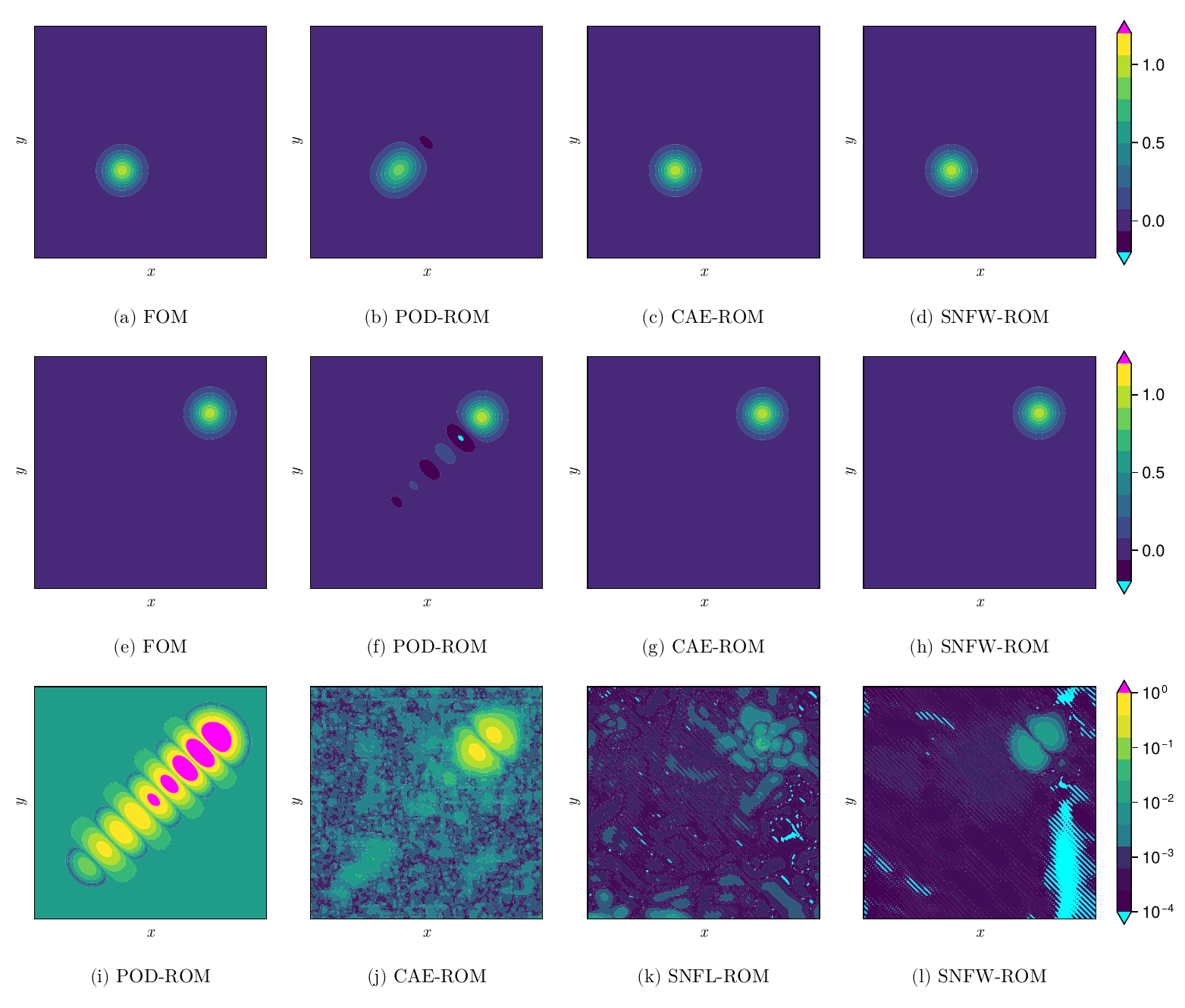}
    \vspace{-2em}
    \caption{
    2D scalar advection test case.
    Top row: prediction at $t = 1$.
    Middle row:
    prediction at $t=4$.
    Bottom row:
    error in prediction $\epsilon(\vect{x}, t)$ at $t=4$.
    The nonlinear ROMs match the FOM solution at both times, whereas POD-ROM produces large numerical oscillations.
    At $t=4$, SNFL-ROM and SNFW-ROM produce much smaller error values than POD-ROM and CAE-ROM.
    The predictions from SNFL-ROM at $t=1, \, 4$, omitted due to space constraints, are indistinguishable from the FOM solution as indicated by the low error values in (k) and in \autoref{fig:exp2f2}.
    }
    \label{fig:exp2f3}
\end{figure}

\begin{figure}[!ht]
    \centering
    \includegraphics[width=0.6\textwidth]{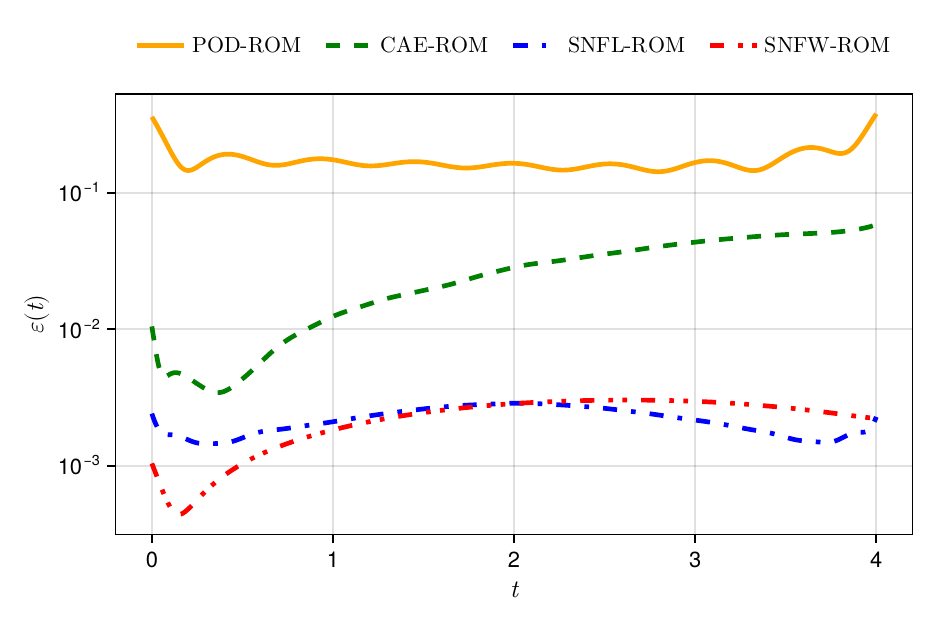}
    \vspace{-1em}
    \caption{
    2D scalar advection test case:
    temporal variation of error in ROM predictions. POD-ROM has the largest error at all time instances.
    The error in CAE-ROM grows with time, while SNFL-ROM and SNFW-ROM produce low errors that do not increase with time.
    }
    \label{fig:exp2f2}
\end{figure}

In this test case, we demonstrate the applicability of our model reduction approach for a 2D advection problem given by the equation
\eqn{
    \label{eqn:advection2D}
    \frac{\p u}{\p t} + \vect{c}\cdot\grad u = 0,
}
where $\vect{c} = \mat{0.25 &\hspace{-0.5em} 0.25}^T$ is the advection velocity.
The solution is initialized with
\begin{equation}
    u_0(\vect{x}) = \exp\left\{
        \frac{-\norm{\vect{x} - \vect{\mu}_0}_2^2}{2\sigma_0^2}
        \right\}
    ,
\end{equation}
where $\vect{\mu}_0 = \mat{-0.5 &\hspace{-0.5em} -0.5}^T$ and $\sigma_0 = 0.1$.

The solution prediction at different time instances is shown in \autoref{fig:exp2f3}.
We observe oscillations in POD-ROM predictions even at an early time of $t = 1$.
These oscillations grow significantly and distort the solution by time $t = 4$.
The nonlinear ROMs, which include CAE-ROM, and SNFW-ROM, yield accurate results for both instances that match the FOM solution well.
The final row of \autoref{fig:exp2f3} presents the distribution of $\epsilon(\vect{x}, t)$ at the final time step for the four ROMs considered in this study.
We notice that POD-ROM oscillations have grown larger in magnitude than the FOM solution.
CAE-ROM has markedly lower errors in large parts of the domain.
However, this error is roughly $10\%$ near the solution peak.
SNFL-ROM has a much smaller overall error, whereas 
SNFW-ROM has the lowest error among all models with $\epsilon(\vect{x}, t) < 1\%$ everywhere in $\Omega$.
A more quantitative comparison of accuracy can be observed by assessing the error metric defined in \autoref{fig:exp2f2}.

The temporal variation of $\varepsilon(t)$ is shown in \autoref{fig:exp2f2}. Consistently high errors are observed for POD-ROM compared to nonlinear ROMs.
CAE-ROM exhibits comparatively lower error for the initial time $t < 0.5$. However, this error grows in time to about $10\%$ of the solution magnitude.
On the other hand, both SNFL-ROM and SNFW-ROM produce consistently low errors that do not grow in time, with SNFW-ROM exhibiting the lowest errors.
With nearly constant errors, SNFL-ROM and SNFW-ROM are better suited for longer time dynamics predictions compared to POD-ROM and CAE-ROM.
These results highlight the robustness of smooth neural fields in yielding consistently accurate representation for 2D advection-dominated problems, which have proven challenging for other ROM approaches as observed from the results.

\begin{table}[!ht]
    \centering
    \caption{
        \noteA{
        2D scalar advection test case:
        speedup over FOM computation and relative error by evaluating SNFW-ROM for different choices of time-step size and number of hyper-reduction points.
        The entries in  each cell are the (top) speed-up with respect to FOM, (middle) the percent relative error at the final timestep $\varepsilon~(t=T)$, and (bottom) GPU memory utilized in the calculation.
        A single FOM evaluation (with $16,384$ grid points and DoFs) for this case takes $0.44\mbox{s}$ of wall time and allocates $1.60\mbox{GiB}$ of GPU memory.
        In contrast, our fastest ROM evaluation takes $0.0619\mbox{s}$ of wall time and allocates only $185\mbox{MiB}$ of GPU memory.
        }
    }
    \vspace{-0em}
    \begin{tabular}{c|*{6}{c|} }
    
    %&&\multicolumn{5}{c}{Number of hyper-reduction points $(\abs{X_\text{proj}})$}
    \multicolumn{7}{c}{\hspace{8em} Number of hyper-reduction points $(\abs{X_\text{proj}})$}
    \\ \cline{2-7}

        &
        %\shortstack{\\Speedup \\ Error ($e$)}
        &
        $16,384$ &
        $4,096$ &
        $1,024$ &
        $256$ &
        $64$
        \\ \cline{2-7}

        \parbox[t]{4mm}{\multirow{4}{*}{\rotatebox[origin=c]{90}{Time-step size $(\Delta t)$}}}
        & \shortstack{\\ \\ $1\Delta t_0$ \\ \textcolor{white}{0}}
        & \shortstack{\\ $0.0465 \times$ \\ $0.296\%$ \\ $355  \mbox{GiB}$}
        & \shortstack{\\ $0.160  \times$ \\ $0.296\%$ \\ $89.2 \mbox{GiB}$}
        & \shortstack{\\ $0.400  \times$ \\ $0.295\%$ \\ $22.6 \mbox{GiB}$}
        & \shortstack{\\ $0.637  \times$ \\ $0.271\%$ \\ $5.95 \mbox{GiB}$}
        & \shortstack{\\ $0.650  \times$ \\ $0.237\%$ \\ $1.79 \mbox{GiB}$}
        \\ \cline{2-7}                                 

        & \shortstack{\\ \\ $2\Delta t_0$ \\ \textcolor{white}{0}}
        & \shortstack{\\ $0.0939 \times$ \\ $0.495\%$ \\ $178  \mbox{GiB}$}
        & \shortstack{\\ $0.327  \times$ \\ $0.495\%$ \\ $44.6 \mbox{GiB}$}
        & \shortstack{\\ $0.818  \times$ \\ $0.494\%$ \\ $11.3 \mbox{GiB}$}
        & \shortstack{\\ $1.313  \times$ \\ $0.465\%$ \\ $2.98 \mbox{GiB}$}
        & \shortstack{\\ $1.338  \times$ \\ $0.393\%$ \\ $911  \mbox{MiB}$}
        \\ \cline{2-7}                                 
                                                       
        & \shortstack{\\ \\ $5\Delta t_0$ \\ \textcolor{white}{0}}
        & \shortstack{\\ $0.238 \times$ \\ $1.12 \%$ \\ $77.4 \mbox{GiB}$}
        & \shortstack{\\ $0.828 \times$ \\ $1.12 \%$ \\ $17.9 \mbox{GiB}$}
        & \shortstack{\\ $2.41  \times$ \\ $1.12 \%$ \\ $4.54 \mbox{GiB}$}
        & \shortstack{\\ $3.34  \times$ \\ $1.09 \%$ \\ $1.19 \mbox{GiB}$}
        & \shortstack{\\ $3.39  \times$ \\ $0.926\%$ \\ $368  \mbox{MiB}$}
        \\ \cline{2-7}                                 
                                                       
        & \shortstack{\\ \\ $10\Delta t_0$ \\ \textcolor{white}{0}}
        & \shortstack{\\ $0.485 \times$ \\ $2.19\%$  \\ $35.8 \mbox{GiB}$}
        & \shortstack{\\ $1.72  \times$ \\ $2.19\%$  \\ $8.99 \mbox{GiB}$}
        & \shortstack{\\ $4.98  \times$ \\ $2.19\%$  \\ $2.28 \mbox{GiB}$}
        & \shortstack{\\ $6.71  \times$ \\ $2.15\%$  \\ $614  \mbox{MiB}$}
        & \shortstack{\\ $6.80  \times$ \\ $1.84\%$  \\ $185  \mbox{MiB}$}
        \\ \cline{2-7}

    \end{tabular}

    \label{tab:exp2hyper}
\end{table}

\noteA{
To assess the effect of time-step size $\Delta t$ and choice of hyper-reduction points $X_\text{proj}$,
we evaluate SNF-ROM at different choices of $\Delta t$ and $\abs{X_\text{proj}}$.
\autoref{tab:exp2hyper} presents speed-up, relative error, and GPU memory allocations as a function of time-step size $\Delta t$ and number of hyper-reduction points $\abs{X_\text{proj}}$.
The speed-up is computed with respect to the FOM computation which employs a Fourier spectral spatial discretization and an adaptive, strong stability preserving Runge-Kutta time-integrator.
Both FOM and ROM computations are carried out on a single GPU device with $11\mbox{GiB}$ of memory.
GPUs are massively parallel machines that can perform fast vector operations on large arrays.
However, GPUs are memory-limited devices and for large array sizes, may run out of memory.
As such, the wall time can be dominated by garbage collection in dynamically typed programming languages.
}

\noteA{
In this experiment, the FOM computation allocates $1.60\mbox{GiB}$ of GPU memory to solve the advection problem over $16,384$ grid points.
Such an allocation can fit comfortably in the VRAM of our GPU device.
Thus, the FOM computation is executed efficiently in $0.44\mbox{s}$ of wall time.
We note that the GPU is under-utilized in this experiment as the FOM is relatively small.
As such, we do not see an appreciable speed-up when $\abs{X_\text{proj}}$ is large.
SNF-ROM allocates $100\times$ more memory than the FOM when $\abs{X_\text{FOM}}=16,384$.
The large allocations here are due to the evaluation of the MLP $g_\theta$ on a large batch of points $X_\text{proj}$.
This leads to frequent calls to the garbage collector during ROM evaluation, leading to a much longer wall time.
}

\noteA{
The memory allocations diminish proportionally as $X_\text{proj}$ is reduced and $\Delta t$ is increased.
This leads to a consistent reduction in wall-time till $\abs{X_\text{proj}} = 256$.
For smaller array sizes in this range, the wall time is dominated by the latency in communication between the host CPU and the GPU device.
For that reason, we do not see a substantial increase in speed-up between $\abs{X_\text{proj}} = 256$ and $\abs{X_\text{proj}} = 64$.
}

\noteA{
We consider the variation of relative error with $\Delta t$ and $\abs{X_\text{proj}}$.
In this experiment, we note that relative error does not significantly vary with $\abs{X_\text{proj}}$.
However, the error increases substantially as we increase the time step size $\Delta t$.
Note that in this work, $X_\text{proj}$ is obtained by uniformly coarsening the FOM grid.
A more sophisticated strategy for choosing $X_\text{proj}$ may further lower error for the same $\abs{X_\text{proj}}$.
Finally, we note the following oddity: the relative error goes down as we reduce $\abs{X_\text{proj}}$.
Further investigation is necessary to ascertain the reason behind this trend.
}

%====================================================%
\subsection{1D viscous Burgers test case}
\label{subsec:exp3}
%====================================================%

\begin{figure}[t]
    \centering
    \includegraphics[width=\textwidth]{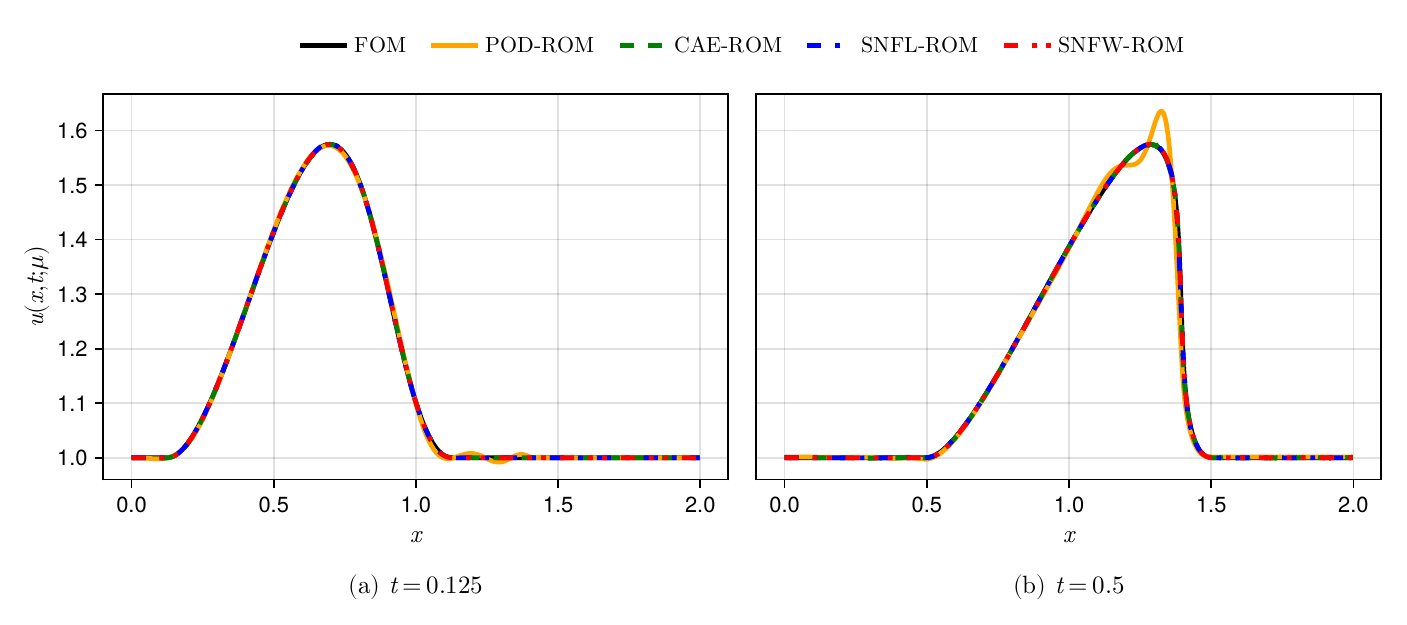}
    \vspace{-2em}
    \caption{
    1D viscous Burgers test case:
    comparison of ROM predictions with the FOM solution at $\mu = 0.575$.
    The predicted solutions for CAE-ROM, SNFL-ROM, and SNFW-ROM agree well with the FOM solution, while POD-ROM produces spurious oscillations near the shock.
    }
    \label{fig:exp3f1}
\end{figure}

We consider 1D viscous Burgers equation,
\begin{equation}
    \frac{\p u}{\p t} + u \frac{\p u}{\p x} = \nu \frac{\p^2 u}{\p x^2},
\end{equation}
where the viscosity $\nu = 10^{-4}$ resulting in a Reynolds number of $10,000$.
The PDE is initialized with a sinusoidal solution profile,
\begin{equation}
     u_0(x; \mu) = \begin{cases}
         1 + \frac{\mu}{2}\left(1 + \sin\left(2\pi x - \frac{\pi}{2} \right) \right) & x \in [0, 1]\\
         0 & \text{otherwise}
     \end{cases},
\end{equation}
parameterized by a scalar $\mu \in \R_+$.
The solution to this equation involves the transport of a wavefront that eventually forms a shock.
Accurate resolution of this shock typically requires a large number of linear modes from POD, and a key goal of nonlinear ROMs is to handle such physics with smaller reduced space representation.
This test case is also used to demonstrate the applicability of proposed ROM approaches for parameter space exploration problems.
In particular, FOM simulations with $\mu_\text{train} = \{0.5, 0.55, 0.6\}$ are used to train the ROM models.
The ROMs are then evaluated for $\mu_\text{test} = \{ 0.525, 0.575, 0.625\}$.
These tests enable the comparison of ROM approaches for both parameter space interpolation ($\mu \in \{0.525, 0.575\}$) and extrapolation ($\mu = 0.625$).

The comparison of different ROMs with FOM at two different time instances is shown in \autoref{fig:exp3f1} for $\mu = 0.575$.
All ROMs agree well with the FOM at $t = 0.25$s, though POD-ROM exhibits small oscillations on the downstream end of the wave. 
These oscillations grow significantly and exhibit a large deviation from the FOM result near the shock at $t = 0.5$.
All nonlinear model reduction approaches exhibit high accuracy and adequately capture the shock. 

\begin{figure}[t]
    \centering
    \includegraphics[width=\textwidth]{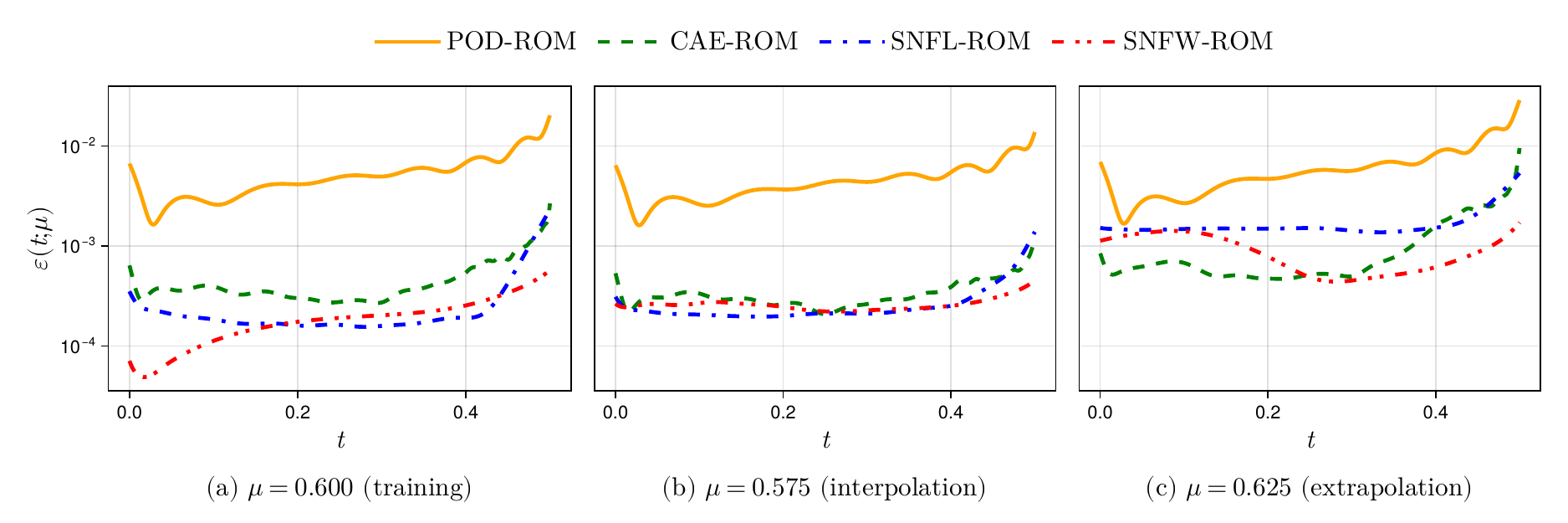}
    \vspace{-2em}
    \caption{
    1D viscous Burgers test case:
    variation of error in ROM prediction against time.
    All ROMs experience a rise in error with time on the extrapolation case ($\mu = 0.625$). POD-ROM has the largest error. CAE-ROM and SNFL-ROM have similar error values, whereas SNFW-ROM performs the best for all cases.
    }
    \label{fig:exp3p2}
\end{figure}

%start new paragraph here.
A more detailed comparison of accuracy involves assessing the evolution of error in time for both training and validation datasets as shown in \autoref{fig:exp3p2}.
% Training
The three nonlinear ROMs perform well on the training case (\autoref{fig:exp3p2}(a): $\mu = 0.600$), whereas POD-ROM exhibits large errors.
This was expected due to its inability to capture the shock in \autoref{fig:exp3f1}.
While all models experience growth in error with time, only SNFW maintains $<0.1\%$ error in $\varepsilon(t; \vect{\mu})$ for all time.
% Interpolation
For all models, the errors appear to be consistent between the training case and the interpolation case, indicating that their performance does not deteriorate significantly within the parameter space.

For the interpolation case (\autoref{fig:exp3p2}(b): $\mu = 0.575$), SNFW-ROM yields the lowest errors with time, whereas CAE-ROM and SNFL-ROM yield slightly larger errors.
% Extrapolation
The errors on the extrapolation case (\autoref{fig:exp3p2}(c): $\mu = 0.625$) are markedly larger for all nonlinear ROMs, though POD-ROM still produces the greatest error values.
The error in SNFL-ROM is consistent at $0.1\%$ until $t = 0.4$, after which it increases to $1\%$.
Similarly, the error in CAE-ROM remains consistent at a low value close to the end of the simulation, when it shoots up to $1\%$.
Although SNFW-ROM experiences a rise in error close to $t=0.5$, the ROM maintains $\sim0.1\%$ error at all times.

\begin{figure}[!ht]
    \centering
    \includegraphics[width=\textwidth]{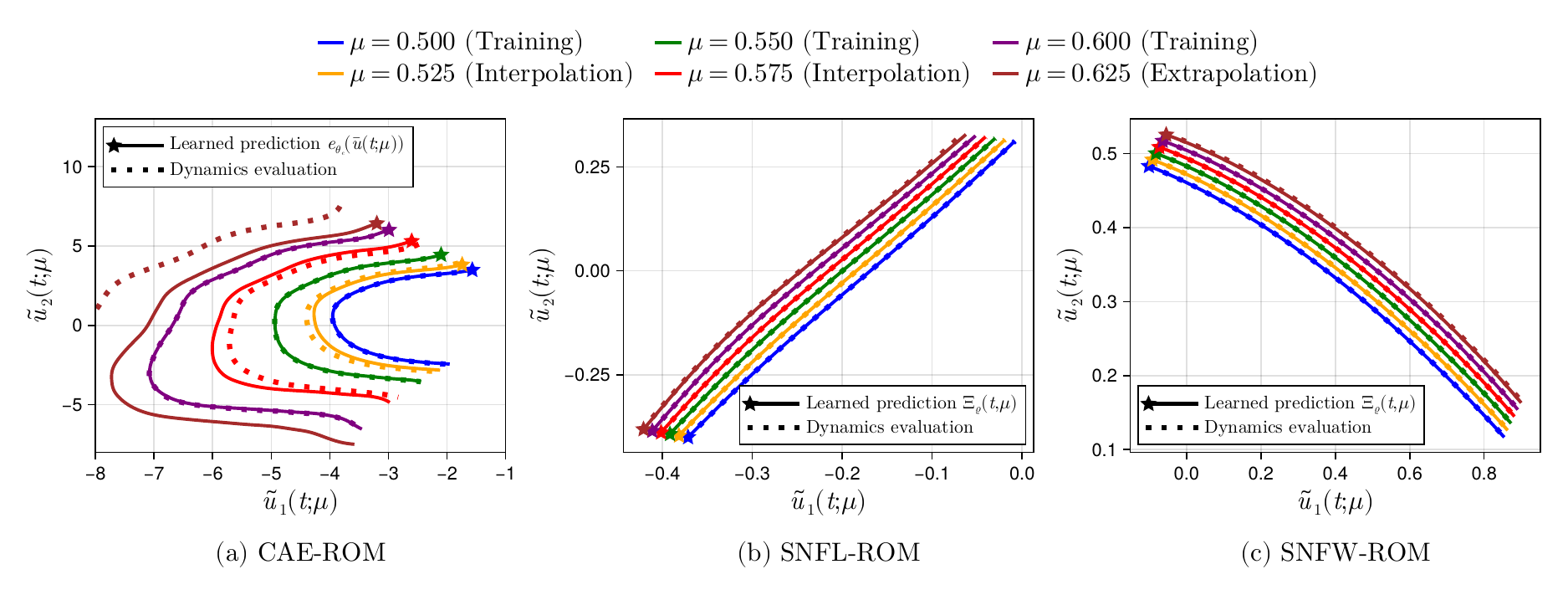}
    \vspace{-2em}
    \caption{
    1D viscous Burgers test case:
    distribution of ROM state vectors
    $\tilde{u}(t; {\mu}) = \mat{\tilde{u}_1(t; {\mu}) &\hspace{-0.5em} \tilde{u}_2(t; {\mu})}^T$.
    The stars mark the beginning of the learned trajectories at $t=0$.
    %
    %The solid lines are the prediction of $\tilde{u}(t; \vect{\mu})$ over $t\in[0, T]$ learned during training with the star marking $\tilde{u}$ at time zero.
    %The dotted lines correspond to $\tilde{u}(t; \vect{\mu})$ obtained from the dynamics evaluation.
    In comparison to CAE-ROM, the SNFL-ROM and SNFW-ROM produce smooth, evenly-spaced learned trajectories that match with $\tilde{u}(t, \vect{\mu})$ from the dynamics evaluation.
    These results facilitate meaningful and robust interpolation and extrapolation in the parameter space.
    }
    \label{fig:exp3f}
\end{figure}

To investigate the source of the error spikes near $t=0.5$, we examine the distribution of ROM state vectors $\tilde{u}$ for the nonlinear ROMs in \autoref{fig:exp3f}.
% CAE shape
The trajectories learned by CAE-ROM (\autoref{fig:exp3f}(a)) do not appear smooth and evenly spaced.
% CAE interpolation
For the interpolation case ($\mu = 0.575$), the ROM state trajectories from the dynamics evaluation deviate from the learned trajectory, although this does not correspond to a marked rise in error.

% CAE extrapolation
In the extrapolation case ($\mu = 0.625$), the ROM state trajectories from the dynamics evaluation of CAE-ROM (\autoref{fig:exp3f}(a)) deviate significantly from the learned trajectory for all time, although the error is low for $t < 0.4$.
% nonlinear
This observation highlights that the deviation of ROM state trajectories from the dynamics evaluation away from the learned trajectory does not necessarily imply poor performance.
As nonlinear ROMs parameterized by neural network functions may have multiple ROM state trajectories that yield low errors,
the behavior of the dynamics evaluation away from the learned trajectory is unknown.

% CAE extrapolation
For the extrapolation case ($\mu = 0.625$), the starting point for the dynamics evaluation for CAE-ROM (\autoref{fig:exp3f}(a)) has moved away from the encoder prediction by Gauss-Newton iteration in the manifold projection step (\autoref{subsec:manifold-projection}).
We found that the Gauss-Newton step in manifold projection (\autoref{eqn:CAE-proj}) significantly improved the accuracy of CAE-ROM over initializing the dynamics evolution with encoder prediction at $t=0$.
Although the ROM state trajectories from the dynamics evaluation remain away from the learned trajectory for all time, the error remains small till $t=0.4$.
The large spike in error at $t>0.4$ illustrates that the dynamics evaluation is unreliable (away from the learned trajectory) and may produce inaccurate results.

% SNF shape
We now investigate the ROM trajectories of SNFL-ROM and SNFW-ROM in \autoref{fig:exp3f}(b), \autoref{fig:exp3f}(c) respectively.
The manifold constraint applied to SNFL-ROM and SNFW-ROM in \autoref{subsec:manifold-constraint} ensures that the trajectories are simple and evenly spaced.
We note that for SNFL-ROM and SNFW-ROM, the ROM trajectories from the dynamics evaluation follow the learned trajectories for all $\mu$.
In the extrapolation case ($\mu = 0.625$),
we observe a deviation between the learned trajectory and the dynamics evaluation for SNFL-ROM and SNFW-ROM near the end of the simulation.
This deviation corresponds to the spike in error in both at $t=0.4$ in that case, although further investigation is needed to ascertain causality.
% conclude this section
This experiment highlights that SNF models are ideally suited for parametric space exploration problems and exhibit higher accuracy than other model reduction approaches considered in this article for both interpolation and extrapolation scenarios.

%====================================================%
\subsection{2D viscous Burgers test case}
\label{subsec:exp4}
%====================================================%

\begin{figure}[ht!]
    \centering
    \includegraphics[width=\linewidth]{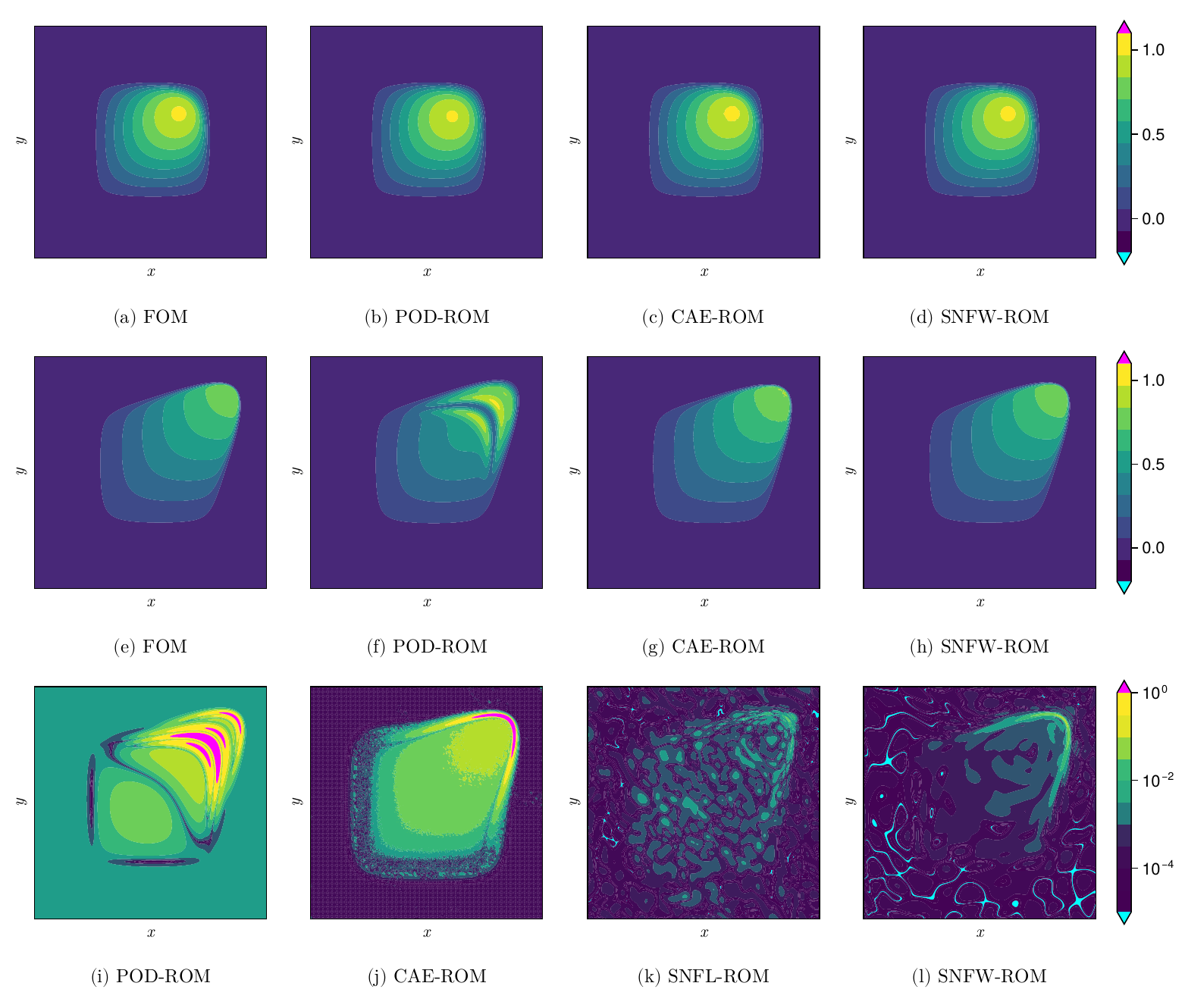}
    \vspace{-2em}
    \caption{
    2D viscous Burgers test case evaluated at $\mu = 1.00$ (interpolation):
    Top row:
    prediction at $t = 0.125$.
    Middle row:
    prediction at $t=0.5$.
    Bottom row:
    error $\epsilon(\vect{x}, t; \mu)$ in prediction for ROM based at $t=0.5$.
    The nonlinear ROMs match the FOM solution at both times, while POD-ROM produces large oscillations.
    A quantitative error assessment indicates that SNFL-ROM and SNFW-ROM significantly outperform POD-ROM and CAE-ROM.
    The predictions from SNFL-ROM at $t=0.125, \, 0.5$, omitted due to space constraints, are indistinguishable from the FOM solution as indicated by the low error values in (k) and in \autoref{fig:exp4p2}.
    }
    \label{fig:exp4f3}
\end{figure}

\begin{figure}[ht!]
    \centering
    \includegraphics[width=\textwidth]{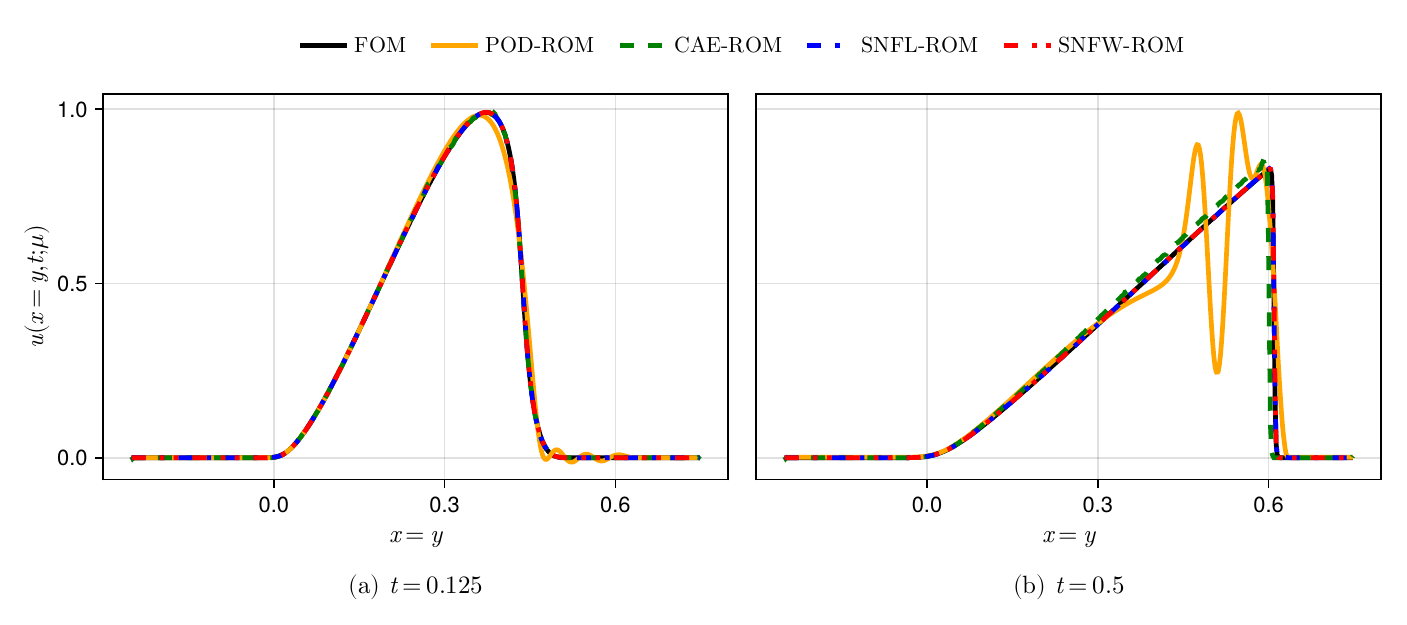}
    \vspace{-2em}
    \caption{
    2D viscous Burgers test case evalauted at $\mu = 1.00$ (interpolation):
    comparison of ROM predictions with FOM solution along the slice $y = x$.
    CAE-ROM, SNFL-ROM and SNFW-ROM capture the steep shock at $t=0.5$, whereas POD-ROM suffers from large oscillations near the shock.
    }
    \label{fig:exp4p1}
\end{figure}

\begin{figure}[ht!]
    \centering
    \includegraphics[width=0.6\textwidth]{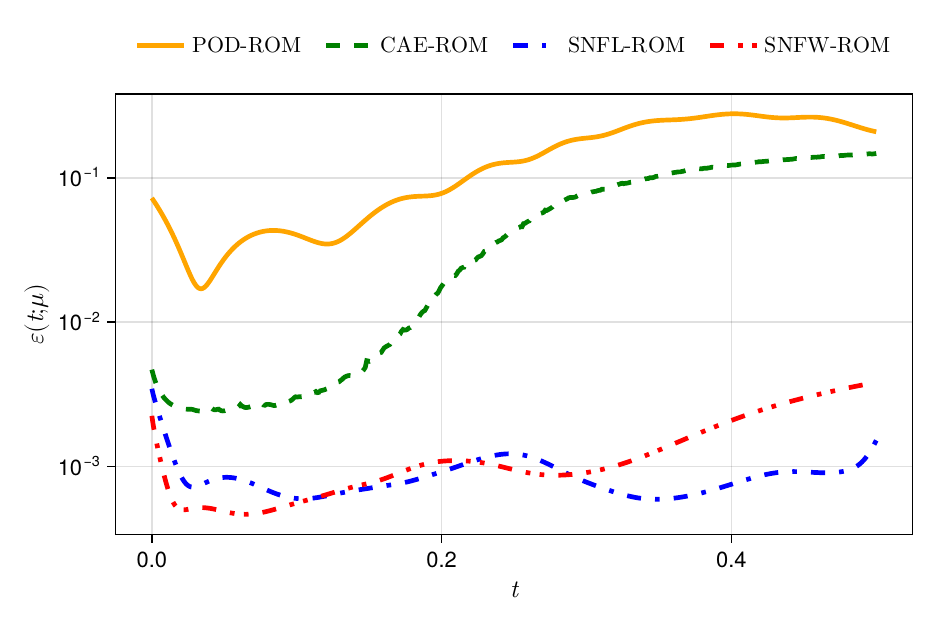}
    \vspace{-1em}
    \caption{
    2D viscous Burgers test case evaluated at $\mu = 1.00$ (interpolation):
    temporal variation of error in ROM solutions.
    SNFL-ROM and SNFW-ROM produce low errors that do not grow with time, whereas errors for POD-ROM and CAE-ROM are high and grow with time.
    }
    \label{fig:exp4p2}
\end{figure}

To demonstrate the applicability of nonlinear model reduction for 2D problems with a shock front, we consider the 2D Burgers equation
\eqn{
    \label{eqn:burg2D}
    \frac{\p \vect{u}}{\p t} + \vect{u} \cdot \nabla \vect{u} = \nu \Delta \vect{u},
}
where $\nu = 10^{-3}$ is the viscosity resulting in a Reynolds number of $1,000$.
Similar to the 1D Burgers test case, the solution propagates in time leading to a shock at $t = 0.5$.
The simulation is initialized to
$\vect{u}_0(x, y) = \mat{u_0(x, y) &\hspace{-0.5em} u_0(x, y)}^T$
where the scalar function $u_0$ is selected to be
\begin{equation}
     u_0(x, y; \mu) = \begin{cases}
         \mu\sin\left(2\pi x\right)\sin\left(2\pi y\right) & x, y \in [0, 0.5]\\
         0 & \text{otherwise}
     \end{cases}
\end{equation}
which is parameterized by a scalar $\mu \in \R_+$.
The solution to this problem involves propagation of a wavefront that eventually forms a shock.
The $x$ and $y$ components of $\vect{u}$ will be equal for all times as the problem is symmetric.
Although the ROM models solve for both components, we restrict our analysis in \autoref{fig:exp4f3} and \autoref{fig:exp4p1} to the first component of the solution vector without loss of generality.
We train the ROM models on FOM simulations with
$\mu_\text{train} = \{0.900, 0.933, 0966, 1.033, 1.066, 1.100\}$,
and evaluate for $\mu_\text{test} = \{1.000\}$.
This test enables the comparison of ROM approaches for parameter space interpolation.
The results discussed hereafter are for the interpolation case $\mu = 1.00$.

The solution at multiple time instances for different ROMs is shown in \autoref{fig:exp4f3}.
Similarly to the 1D burgers test case, we observe that POD-ROM exhibits large oscillations and deviation from the FOM results at later times.
All nonlinear model reduction approaches capture the shock well and exhibit much better accuracy than POD-ROM.
A qualitative assessment of $\epsilon(\vect{x}, t)$ indicates that CAE-ROM exhibits large errors near the shock front.
SNFL-ROM and SNFW-ROM produce much smaller errors around the shock, as well as further away, with SNFW-ROM producing $\epsilon\leq 0.1\%$ errors everywhere at $t = 0.5  $.

This observation is more evident from results in \autoref{fig:exp4p1}, which show ROM predictions along the diagonal slice $y = x$.
We observe that POD-ROM exhibits a much higher error throughout the domain than nonlinear ROMs, which produce much more accurate results.
We also observe that CAE-ROM does not adequately capture the shock at $t=0.5$, whereas SNFL-ROM and SNFW-ROM are indistinguishable from the FOM solution.
The temporal evolution of error for different ROMs is shown in \autoref{fig:exp4p2}.
These results indicate that POD-ROM and CAE-ROM exhibit large errors that grow with time.
Conversely, SNFL-ROM and SNFW-ROM experience a decrease in error at an early time, after which the error in SNFW-ROM rises steadily till the final time step, whereas the error in SNFL-ROM stabilizes to $\sim0.1\%$.
These results highlight the suitability of smooth neural fields in capturing the physics of this 2D flow with a shock wave accurately and efficiently. 

\begin{table}[t]
    \centering
    \caption{
        \noteA{
        2D viscous Burgers test case evaluated at $\mu = 1.00$ (interpolation):
        we present speedup over FOM computation and relative error for different choices of time-step size and number of hyper-reduction points.
        The entries in  each cell are (top) the speed-up with respect to FOM, (middle) the percent relative error at the final timestep $\varepsilon~(t=T; \mu)$, and (bottom) the GPU memory utilized in the calculation.
        A single FOM evaluation (with $262\mbox{k}$ points and $524\mbox{k}$ DoFs) for this case takes $13.44\mbox{s}$ of wall time and allocates $640\mbox{GiB}$ of GPU memory.
        In contrast, our fastest ROM evaluation takes $0.0684\mbox{s}$ of wall time and allocates only $261\mbox{MiB}$ of GPU memory.
        }
    }
    \vspace{-0em}
    \begin{tabular}{c|*{6}{c|} }
    
    %&&\multicolumn{5}{c}{Number of hyper-reduction points $(\abs{X_\text{proj}})$}
    \multicolumn{7}{c}{\hspace{8em} Number of hyper-reduction points $(\abs{X_\text{proj}})$}
    \\ \cline{2-7}

        &
        % \shortstack{\\Speedup \\ Error ($e$)} &
        &
        $16,384$ &
        $4,096$ &
        $1,024$ &
        $256$ &
        $64$
        \\ \cline{2-7}

        \parbox[t]{4mm}{\multirow{4}{*}{\rotatebox[origin=c]{90}{Time-step size $(\Delta t)$}}}
        & \shortstack{\\ \\ $1\Delta t_0$ \\ \textcolor{white}{0}}
        & \shortstack{\\ $1.92 \times$ \\ $0.186\%$ \\ $546  \mbox{GiB}$}
        & \shortstack{\\ $4.39 \times$ \\ $0.231\%$ \\ $137  \mbox{GiB}$}
        & \shortstack{\\ $11.8 \times$ \\ $0.428\%$ \\ $34.5 \mbox{GiB}$}
        & \shortstack{\\ $18.3 \times$ \\ $0.342\%$ \\ $8.93 \mbox{GiB}$}
        & \shortstack{\\ $19.1 \times$ \\ $1.16 \%$ \\ $2.53 \mbox{GiB}$}
        \\ \cline{2-7}  

        & \shortstack{\\ \\ $2\Delta t_0$ \\ \textcolor{white}{0}}
        & \shortstack{\\ $2.42 \times$ \\ $0.171 \%$ \\ $273  \mbox{GiB}$}
        & \shortstack{\\ $8.85 \times$ \\ $0.211 \%$ \\ $68.5 \mbox{GiB}$}
        & \shortstack{\\ $23.7 \times$ \\ $0.469 \%$ \\ $17.3 \mbox{GiB}$}
        & \shortstack{\\ $38.9 \times$ \\ $0.361 \%$ \\ $4.47 \mbox{GiB}$}
        & \shortstack{\\ $38.7 \times$ \\ $1.16  \%$ \\ $1.27 \mbox{GiB}$}
        \\ \cline{2-7}  
  
        & \shortstack{\\ \\ $5\Delta t_0$ \\ \textcolor{white}{0}}
        & \shortstack{\\ $6.08 \times$ \\ $0.168\%$ \\ $110  \mbox{GiB}$}
        & \shortstack{\\ $22.3 \times$ \\ $0.185\%$ \\ $27.4 \mbox{GiB}$}
        & \shortstack{\\ $61.7 \times$ \\ $0.588\%$ \\ $6.92 \mbox{GiB}$}
        & \shortstack{\\ $99.9 \times$ \\ $0.706\%$ \\ $1.79 \mbox{GiB}$}
        & \shortstack{\\ $101  \times$ \\ $1.64 \%$ \\ $520  \mbox{MiB}$}
        \\ \cline{2-7}  
  
        & \shortstack{\\ \\ $10\Delta t_0$ \\ \textcolor{white}{0}}
        & \shortstack{\\ $12.2 \times$ \\ $0.212\%$ \\ $54.9 \mbox{GiB}$}
        & \shortstack{\\ $44.5 \times$ \\ $0.302\%$ \\ $13.8 \mbox{GiB}$}
        & \shortstack{\\ $138  \times$ \\ $1.25 \%$ \\ $3.47 \mbox{GiB}$}
        & \shortstack{\\ $199  \times$ \\ $1.09 \%$ \\ $919  \mbox{MiB}$}
        & \shortstack{\\ $199  \times$ \\ $0.369\%$ \\ $261  \mbox{MiB}$}
        \\ \cline{2-7}

    \end{tabular}

    \label{tab:exp4hyper}
\end{table}

\noteA{
In order to demonstrate a speedup over the FOM computation, we evaluate SNFL-ROM model for different choices of time-step size $\Delta t$ and number of hyper-reduction points $\abs{X_\text{proj}}$.
\autoref{tab:exp4hyper} presents speed-up and relative error as a function of $\Delta t$ and $\abs{X_\text{proj}}$.
In this experiment, the FOM calculation allocates a total of $640\mbox{GiB}$ of GPU memory to solve \autoref{eqn:burg2D} over $262\mbox{k}$ points, and takes $13.44\mbox{s}$ of wall time.
}

\noteA{
The first trend we note in \autoref{tab:exp4hyper} is that memory allocation decreases proportionally as we reduce $\abs{X_\text{proj}}$ and increase $\Delta t$.
This is expected as reducing $\abs{X_\text{proj}}$ leads to evaluating $g_\theta$ on smaller batches, and increasing $\Delta t$ leads to fewer calculations altogether.
We note that speed-up increases proportionally with $\Delta t$ for all $\abs{X_\text{proj}}$.
We also note a similar steady rise speed-up with a reduction in $\abs{X_\text{proj}}$ till $\abs{X_\text{proj}} = 256$.
Thereafter, we observe no substantial increase in speed-up as the wall-time is dominated by latency of CPU to GPU communication.
The maximum speed-up achieved in this experiment is $199\times$ against the FOM calculation.
}

\noteA{
In \autoref{tab:exp4hyper}, we note that the relative error increases as we decrease $\abs{X_\text{proj}}$.
This is expected as the dynamics associated with the omitted collocation points is not available to guide the time-integration of the reduced system.
We also note that, as expected, the relative error increases steadily with $\Delta t$.
Both trends in relative error are thwarted by the case with $\Delta t = 10\Delta t_0$ and $X_\text{proj} = 64$, which has a relative error of $0.369\%$.
This might be due to fortuitous cancellation of error introduced by hyper-reduction and time-integration, although further investigation is needed to ascertain the reason for this phenomenon.
}

%====================================================%
\subsection{1D Kuromoto-Sivashinsky test case}
\label{subsec:exp5}
%====================================================%

\begin{figure}[ht!]
    \centering
    \includegraphics[width=\textwidth]{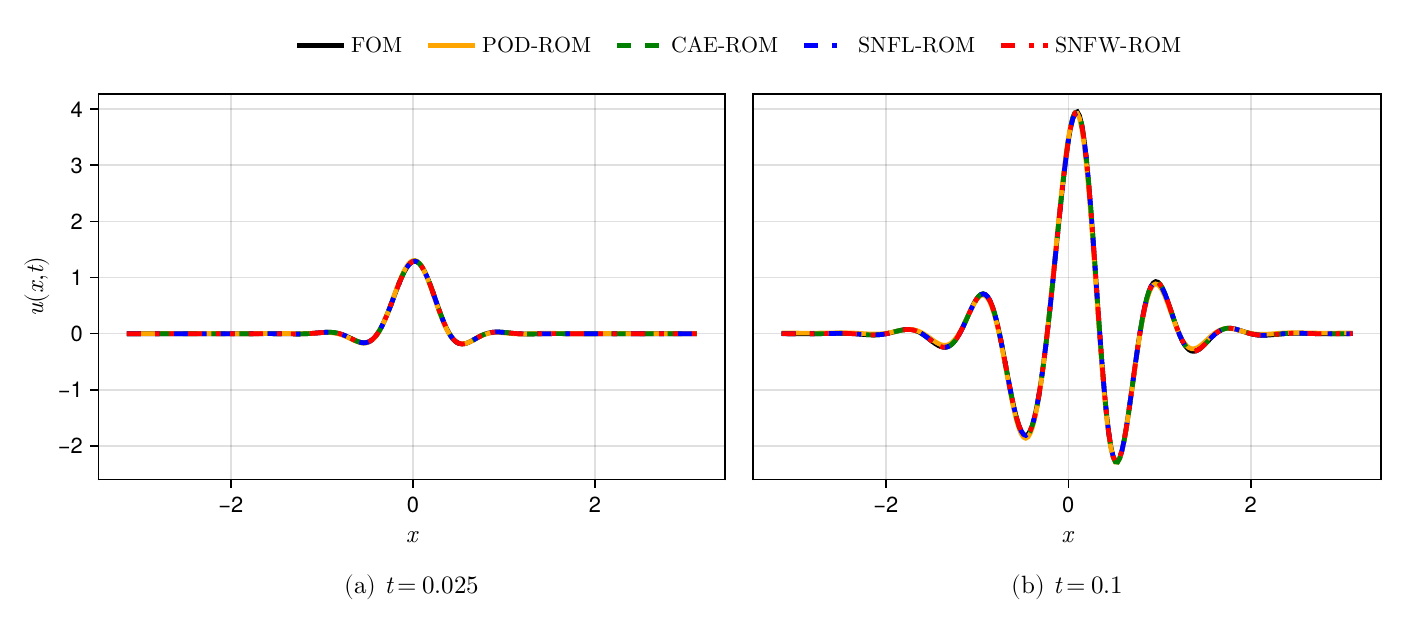}
    \vspace{-2em}
    \caption{
    1D Kuromoto-Sivashinsky test case:
    comparison of ROM predictions with the FOM solution.
    As this problem is highly diffusive ($\nu = 0.01$), all linear and nonlinear ROMs agree with the FOM solution.
    }
    \label{fig:exp5f1}
\end{figure}

For the last test case, we select the Kuramoto-Sivashinsky (KS) equation to compare the performance of ROMs using nonlinear PDEs with higher-order derivatives. This PDE is given as,
\eqn{
    \frac{\p u}{\p t} + u \frac{\p u}{\p x} + \frac{\partial^2 u}{\p x^2} + \nu \frac{\partial^4 u}{\p x^4} = 0,
}
where $\nu = 0.01$ is the viscosity.
Higher-order derivatives in this problem make the dynamics difficult to predict.
In particular, we note that the problem becomes high oscillatory results indicating unstable behavior for large number of POD modes.
As such, we solve this problem with $N_\text{ROM}=2$ for POD-ROM.
In comparison, nonlinear ROMs did not show a significant difference in performance between $N_\text{ROM} = 1, \, 2$.
We present results here for $N_\text{ROM} = 1$ for nonlinear ROMs.
The PDE is initialized with
\begin{equation}
    u_0(x) = \exp\left\{
        \frac{-x^2}{2\sigma_0^2}
        \right\}
    ,
\end{equation}
where $\sigma_0 = 0.2$.
As the solution of this equation is chaotic with a large sensitivity to the initial conditions, we only compare the ROMs at an initial short time window. 

\begin{figure}[t]
    \centering
    \includegraphics[width=\textwidth]{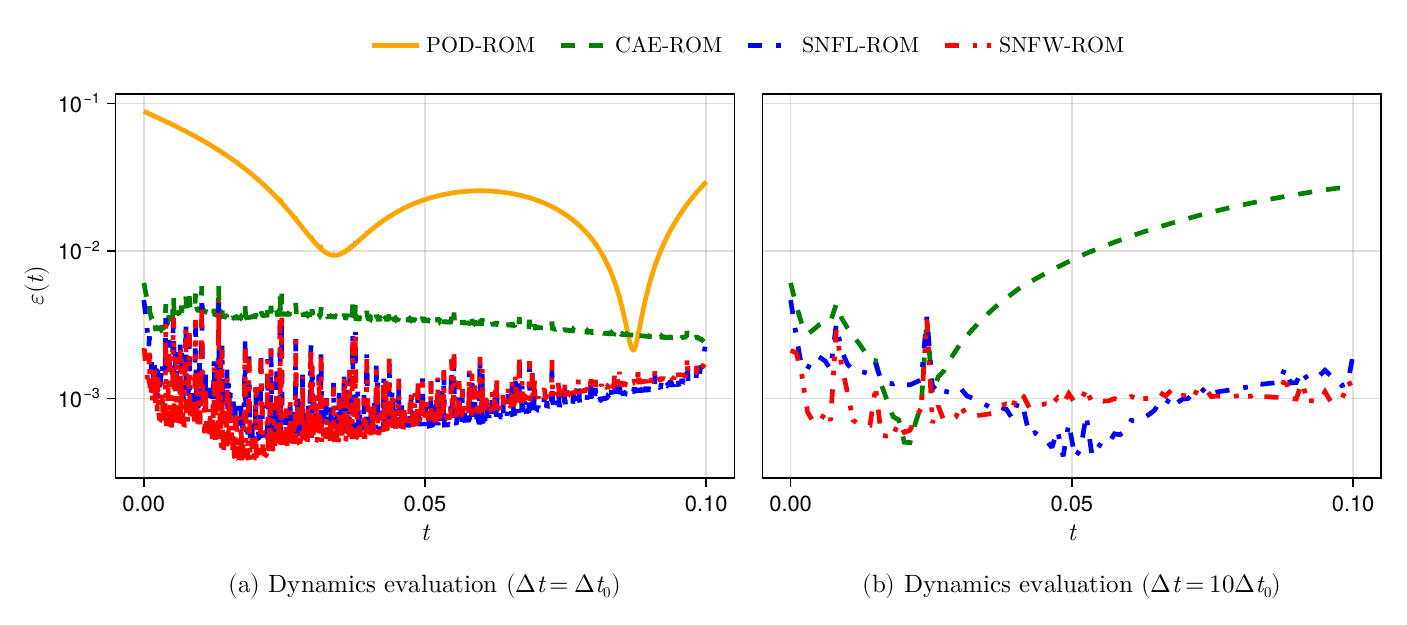}
    \vspace{-1em}
    \caption{
    1D Kuromoto-Sivashinsky test case:
    variation of error in ROM predictions against time.
    Despite large oscillations in error values, the error in nonlinear ROMs remains below $1\%$ at all times.
    SNFL-ROM and SNFW-ROM produce low error values at all times.
    Note that the $y$-axis in this figure has $\log$ scale, which may distort the oscillations in the error curves for the nonlinear ROMs.
    }
    \label{fig:exp5f2}
\end{figure}

A comparison of solutions using different ROMs at two time instances is shown in \autoref{fig:exp5f1}.
The solution grows from a Gaussian initial condition to a complex signal with varying magnitudes.
As the problem is highly diffusive, all ROMs
appear to give similar results that are in good agreement with FOM.
To compare these approaches more closely, we analyze the temporal evolution of the error defined in \autoref{eqn:error_t}.
The temporal evolution of model error for different ROMs is shown in \autoref{fig:exp5f2} which presents the error evolution of ROMs from a dynamics evaluation with $\Delta t = \Delta t_0$ and a dynamics evaluation with $\Delta t = 10\Delta t_0$.
POD-ROMs exhibit the highest error for all instances of time that are tested with both $\Delta t$.
CAE-ROMs exhibit a lower error value that remains consistent over time for $\Delta t = \Delta t_0$.
For $\Delta t = 10\Delta t$, however, the error in POD-ROM rises dramatically with time.
SNF-ROMs exhibit a much smaller error in the initial time window, which does not grow considerably over time.
We observe large oscillations in the error curves for SNFL-ROM and SNFW-ROM in \autoref{fig:exp5f2} whose origin is not fully understood. However, despite these oscillations, the errors are consistently below those for POD-ROM and CAE-ROM indicating superior performance of proposed SNF-ROMs.
Low errors in this numerical experiment indicate that SNF-ROM can adequately capture the dynamics of PDE problems with higher-order derivatives.

%%%%%%%%%%%%%%%%%%%%%%%%%%%%%%%%%%%%%%%%%%%%%%%%%%%%%%%%%%%%%%%%%%%
\section{Conclusions and discussion}
\label{sec:conclusion}
%%%%%%%%%%%%%%%%%%%%%%%%%%%%%%%%%%%%%%%%%%%%%%%%%%%%%%%%%%%%%%%%%%%

% 
SNF-ROM is a machine learning-based nonlinear model order reduction method that predicts dynamics by combining neural field spatial representations from point-cloud datasets with Galerkin projections.
Compared to existing nonlinear ROMs, SNF-ROM presents several advantages in the online stage of dynamics prediction.
SNF-ROM constrains the learned ROM manifold to be smooth and regular, thus allowing for robust online stage evaluations with large time steps.
By supporting AD-based spatial differencing, SNF-ROM drastically simplifies the computation of online dynamics.
Numerical experiments reveal that the dynamics evaluation of SNF-ROM is accurate and that the proposed smoothing methodology is robust, thus eliminating the need for significant hyperparameter tuning.
With a combination of hyper-reduction and larger time-steps, SNF-ROM faster than the full-order model by up to $199\times$.
While offering the key advantages mentioned above, SNF-ROM consistently outperforms other tested ROM approaches across various PDE problems.

% limitations
SNF-ROM is not without limitations.
% training time
For one, the offline training cost is significantly greater than that for a linear POD-ROM.
This higher cost is a significant hurdle in working with nonlinear ROMs, as large training times bottleneck prototyping and experimentation processes.
Future work may consider using modern neural architectures that support fast training \cite{muller_instant_2022}.
Another possible research direction could be to use transformer-based meta-learning approaches to learn network weights \cite{chen_transformers_2022} quickly.
% BC
Many nonlinear ROMs are unable to handle complex and changing boundary conditions;
future work could be directed towards alleviating this limitation.
% performance
%Future work may also consider lowering the cost of the online stage by integrating novel hyper-reduction schemes within SNF-ROM devising hyper-reduction sampling schemes for SNF-ROM.
Further performance gains may be achieved by writing customized GPU kernels for the online stage and employing highly efficient Taylor mode AD \cite{tan2022taylordiff, jax2018github} instead of forward mode AD.
% closing
Nonlinear reduced order modeling is an emerging technology in its nascent stage.
This work has identified and eliminated critical shortcomings in existing approaches. Substantial enhancements must be made for its maturation and practical application.

%%%%%%%%%%%%%%%%%%%%%%%%%%%%%%%%%%%%%%%%%%%%%%%%%%%%%%%%%%%%%%%%%%%
\section*{Acknowledgements}
\label{sec:acknowledgements}
%%%%%%%%%%%%%%%%%%%%%%%%%%%%%%%%%%%%%%%%%%%%%%%%%%%%%%%%%%%%%%%%%%%

The authors would like to acknowledge the support from the National Science Foundation (NSF) grant CMMI-1953323 and PA Manufacturing Fellows Initiative for the funds used towards this project.
The research in this paper was also sponsored by the Army Research Laboratory and was accomplished under Cooperative Agreement Number W911NF-20-2-0175.
The views and conclusions contained in this document are those of the authors and should not be interpreted as representing the official policies, either expressed or implied, of the Army Research Laboratory or the U.S. Government.
The U.S. Government is authorized to reproduce and distribute reprints for Government purposes notwithstanding any copyright notation herein.

%%%%%%%%%%%%%%%%%%%%%%%%%%%%%%%%%%%%%%%%%%%%%%%%%%%%%%%%%%%%%%%%%%%
%%%%%%%%%%%%%%%%%%%%%%%%%%%%%%%%%%%%%%%%%%%%%%%%%%%%%%%%%%%%%%%%%%%
% appendix

\appendix
%%%%%%%%%%%%%%%%%%%%%%%%%%%%%%%%%%%%%%%%%%%%%%%%%%%%%%%%%%%%%%%%%%%
\section{Manifold projection}
\label{sec:appendix-manifold-projection}
%%%%%%%%%%%%%%%%%%%%%%%%%%%%%%%%%%%%%%%%%%%%%%%%%%%%%%%%%%%%%%%%%%%

%==================================%
\subsection{CAE-ROM}
\label{subsec:appendix-cae-rom-projection}
%==================================%

The CAE-ROM manifold projection function $h'_\text{CAE}$ maps FOM state vectors $\dvect{u}\in\R^{N_\text{FOM}}$ to corresponding ROM state vectors ${\tilde{u}}\in\R^{N_\text{ROM}}$
such that $h'_\text{CAE}(\dvect{u})$ reproduces $\dvect{u}$ with $g'_\text{CAE}$ to a reasonable accuracy.
That is,
\eqn{
    \dvect{u} \approx g'_\text{CAE} \circ h'_\text{CAE}(\dvect{u}).
}
Previous CAE-ROM implementations \cite{lee_model_2020, kim_fast_2022} have employed the learned encoder for this task:
\eqn{
    h'_\text{CAE}: \dvect{u} \to e_{\theta_e}(\dvect{u}).
}
In our experiments, we found that the encoder prediction by itself may not be a local optimum solution to the nonlinear system
\eqn{
    \label{eqn:CAE-proj-wo-GN}
    \argmin_{{\tilde{u}}}
    \norm{\dvect{u} - g'_\text{CAE}({\tilde{u}})}^2_2.
}
As such, we solve this nonlinear least squares problem with Gauss-Newton iteration to implement $h'_\text{CAE}$,
\eqn{
    \label{eqn:CAE-proj}
    h'_\text{CAE}(\dvect{u}) = 
    \argmin_{{\tilde{u}}}
    \norm{\dvect{u} - g'_\text{CAE}({\tilde{u}})}^2_2,
}
where the initial guess is the encoder prediction $e_{\theta_e}(\dvect{u})$.
Our experiments indicate that using Gauss-Newton iteration to implement $h'_\text{CAE}$ performed significantly better than simply using the encoder prediction as in \autoref{eqn:CAE-proj-wo-GN}.

%==================================%
\subsection{SNF-ROM}
\label{subsec:appendix-snf-rom-projection}
%==================================%

% SNF-ROM projection
We define the SNF-ROM manifold projection function $h_\theta$.
Given a vector field $\vect{u}:\Omega \to \R^m$, $h_\theta$ seeks ROM state vector
${\tilde{u}}\in\R^{N_\text{ROM}}$
that can reconstruct $\vect{u}$ to a reasonable accuracy. That is,
\eqn{
    \label{eqn:neuralfield-proj-approx}
    \vect{u}(\vect{x}) \approx
    g_\theta \left(\vect{x}, h_\theta(\vect{u}) \right),
    \, \forall \vect{x} \in \Omega.
}
Unlike discretization-dependent ROMs where the projection functions map between finite-dimensional fields, $h_\theta$ maps the continuous fields over $\Omega$ to
$\R^{N_\text{ROM}}$.
We close this problem by selecting a small subset of points
$X_\text{proj} \subset \Omega$ where we attempt to satisfy \autoref{eqn:neuralfield-proj-approx}
\eqn{
    \label{eqn:neuralfield-proj-approx-xproj}
    \vect{u}(\vect{x}) \approx
    g_\theta \left(\vect{x}, h_\theta(\vect{u}) \right),
    \, \forall \vect{x} \in X_\text{proj}.
}
The choice of points in $X_\text{proj}$ is not restricted to $X_\text{FOM}$ and can be sampled anywhere in $\Omega$.
Although only $N_\text{ROM}$ points are needed for solving \autoref{eqn:neuralfield-proj-approx-xproj},
the system is typically over-determined \cite{chen_crom_2023} with $N_\text{ROM} < \abs{X_\text{proj}} << N_\text{FOM}$ where $\abs{X_\text{proj}}$ is the number of points in $X_\text{proj}$.

If the intrinsic coordinates $(t, \vect{\mu})$ of $\vect{u}$ are known, then $\tilde{u}(t; \vect{\mu})$ can be predicted using the intrinsic ROM manifold as $\Xi_\varrho$. That is,
\eqn{
    h_\theta: \vect{u}(t; \vect{\mu}) \to \Xi_\varrho(t, \vect{\mu}).
}
However, our experiments in \autoref{sec:experiments} indicate that $\Xi_\varrho(t, \vect{\mu})$ may not be the local optimum of the system
\eqn{
    \label{eqn:neuralfield-proj}
    \argmin_{\tilde{u}}
    \sum_{\vect{x} \in X_\text{proj}}
    \norm{\vect{u}(\vect{x}) - g_\theta(\vect{x}, \tilde{u})}^2_2.
}
As such, we implement $h_{\theta}$ by solving \autoref{eqn:neuralfield-proj} with Gauss-Newton iteration where the initial guess is given by $\Xi_\varrho(t, \vect{\mu})$.
Our experiments indicate that using Gauss-Newton iteration for manifold projection leads to marginal improvements in accuracy over simply using $\Xi_\varrho(t, \vect{\mu})$.

%%%%%%%%%%%%%%%%%%%%%%%%%%%%%%%%%%%%%%%%%%%%%%%%%%%%%%%%%%%%%%%%%%%
\section{Neural field regularization}
\label{sec:appendix-regularization}
%%%%%%%%%%%%%%%%%%%%%%%%%%%%%%%%%%%%%%%%%%%%%%%%%%%%%%%%%%%%%%%%%%%

Here, we discuss the effectiveness of Lipschitz regularization and weight regularization at smoothing neural field representations.
The purpose of this step is to ensure that spatial derivatives of neural field representations adequately capture the derivatives of the true signal.
A typical neural field representation is not guaranteed to smoothly interpolate the training data.
As such, partial derivatives of neural field ROMs are riddled with numerical artifacts, making them unusable in downstream applications such as time-evolution. In this appendix, we assess the effectiveness of the proposed regularization strategies in \autoref{subsec:neural-field-regularization} at curbing numerical artifacts in spatial derivatives of a neural field representation.

\begin{figure}[t]
    \centering
    \includegraphics[width=0.6\textwidth]{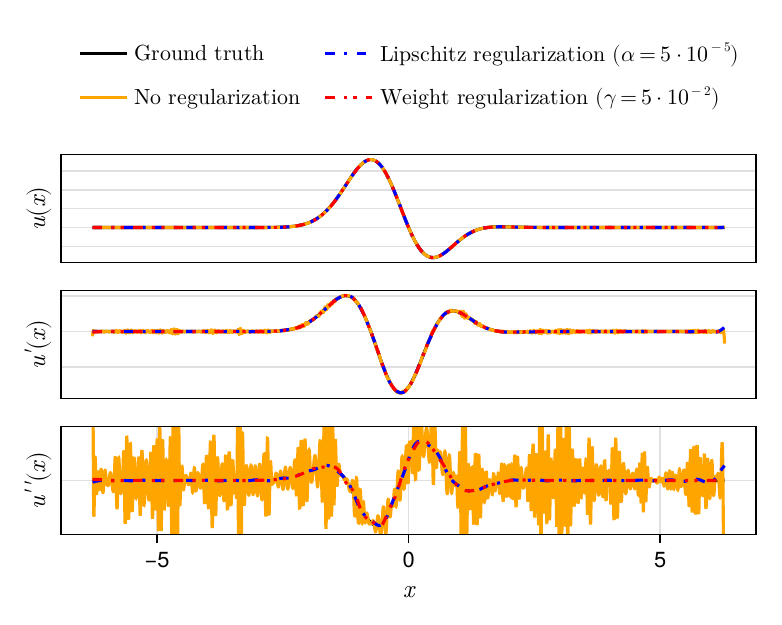}
    \vspace{-5mm}
    \caption{
        We compare the effectiveness of Lipschitz regularization and weight regularization at smoothing neural field representations.
        We train a set of neural networks to regress the function
        $u(x) = (x - \dfrac{\pi}{2}) \sin(x) \exp(-x^2)$ with different regularization strategies (top).
        The first (middle) and second (bottom) derivatives of the neural network representations are computed exactly in automatic differentiation.
        When no regularization (orange line) is applied, the first and second derivatives of the neural field differ from those of the true signal.
        Comparatively, Lipschitz regularization and weight regularization significantly smooth out the noise in spatial derivatives of the neural field representation.
    }
    \label{fig:appendix-regularization}
\end{figure}

In \autoref{fig:appendix-regularization}, we train a set of MLPs to regress a simple $1D$ function with different regularization strategies.
We fix the width of the hidden layer of each MLP to be $64$ and adopt the training strategy and architecture details of $g_\theta$ described in \autoref{sec:appendix-snf-rom-train}.
Spatial derivatives of the neural network representations are then computed exactly with automatic differentiation and compared with the true derivatives of the signal $u(x)$ in \autoref{fig:appendix-regularization}.

\autoref{tab:appendix-regularization} quantifies the mismatch between the derivatives of the neural field representations and the derivatives of the true signal by computing the relative mean-squared error.
We note that all neural fields accurately capture the ground truth function.
However, when no regularization is applied, the first and second derivatives differ from those of the true signal.
This may lead to numerical instabilities in the dynamical evaluation in \autoref{sec:dynamics}.
In contrast, both Lipschitz regularization and weight regularization adequately capture the derivative of the signal.
This is evident in the successful numerical experiment in \autoref{subsec:exp5} where we perform dynamics evaluation on the fourth-order KS problem.

\begin{table}[!ht]
    \caption{
        We compare the effectiveness of Lipschitz regularization and weight regularization at smoothing neural field representations.
        We compare the derivatives of the signal $u(x) = (x - \dfrac{\pi}{2}) \sin(x) \exp(-x^2)$ and those of neural field representations trained to regress $u(x)$ with different regularization strategies.
        The mismatch is quantified with the relative mean squared error.
    }
    \centering
    \begin{tabular}{|c|c|c|c|}
        \hline
         Regularization approach & $u(x)$ & $u'(x)$ & $u''(x)$ \\\hline
         No regularization                                   & $2.75\times10^{-6}$ & $7.06\times10^{-2}$ & $70.1$ \\\hline
         %$L_2$ regularization ($\gamma = 1\cdot10^{-1}$)     & $1.68\cdot10^{-6}$ & $5.06\cdot10^{-5}$ & $1.81\cdot10^{-2}$ \\\hline
         Lipschitz regularization ($\alpha = 5\times10^{-5}$) & $3.68\times10^{-6}$ & $5.28\times10^{-4}$ & $9.72\times10^{-2}$ \\\hline
         Weight regularization ($\gamma = 5\times10^{-2}$)    & $2.10\times10^{-6}$ & $1.01\times10^{-4}$ & $1.76\times10^{-2}$ \\\hline
    \end{tabular}
    \label{tab:appendix-regularization}
\end{table}

%%%%%%%%%%%%%%%%%%%%%%%%%%%%%%%%%%%%%%%%%%%%%%%%%%%%%%%%%%%%%%%%%%%
\section{SNF-ROM: Architecture and training details}
\label{sec:appendix-snf-rom-train}
%%%%%%%%%%%%%%%%%%%%%%%%%%%%%%%%%%%%%%%%%%%%%%%%%%%%%%%%%%%%%%%%%%%

We discuss the implementation details in manifold construction of SNF-ROM.
Given the tuple $\left(\vect{x}, t, \vect{\mu}\right)$, we make a prediction for $\vect{u}(\vect{x}, t; \vect{\mu})$ as follows.
First, we obtain the ROM state vector corresponding to $t$, $\vect{\mu}$ on the intrinsic ROM manifold $\mathcal{\tilde{U}}$ as
${\tilde{u}}(t; \vect{\mu}) = \Xi_\varrho(t, \vect{\mu})$.
Then the query point $\vect{x}$ is concatenated with ${\tilde{u}}(t; \vect{\mu})$.
The concatenated vector is then passed to the MLP, $g_\theta$.
In the backward pass, the gradients are calculated with respect to $\theta$ and $\varrho$.
See \autoref{fig:training-pipeline} for reference.

% decoder network
In all experiments, the network $g_\theta$ has $7$ layers.
The $5$ hidden layers have a fixed width $w$, whereas the sizes of the input and output layers are $d+N_\text{ROM}$ and $m$ respectively.
A \textit{sine} activation function, i.e. $\sigma(x) =  \sin(x)$, is chosen for the ability of sinusoidal networks to fit high-frequency functions \cite{sitzmann_implicit_2020}.
The final layer of the network does not use a bias vector.
% intrinsic manifold parameterization
The network $\Xi_\varrho$ is chosen to be a shallow MLP with width $8$.
The network is purposely chosen to be small so that simple ROM representations are learned.
The \textit{tanh} activation function is chosen to ensure that ROM state vectors are continuous functions of $t$ and $\vect{\mu}$.
Furthermore, we apply $L^2$ regularization with $\gamma = 10^{-3}$ on the parameters $\varrho$ to ensure that a smooth mapping is learned.

% initialization
In $g_\theta$, the weight matrices have been initialized per the SIREN initialization scheme described in \cite{sitzmann_implicit_2020} to ensure a normal distribution of output values following a \textit{sine} activation.
This involves scaling the weight matrices in the network layers by constant $\omega$ during initialization.
For the first layer, $\omega$ is set to $10$ to represent a wide range of frequencies \cite{sitzmann_implicit_2020}, and $\omega = 1$ for the following layers.
The choice of $\omega$ directly affects the range of frequencies the model is able to represent, where low values encourage smoother functions, and larger values encourage high-frequency and finely-detailed functions \cite{wiesner_implicit_2022}.
The final layer is initialized with Xavier initialization \cite{kumar_weight_2017}.
The initial bias vectors are sampled from a uniform random distribution to create a phase shift among the sinusoidals in each layer.
For the network $\Xi_\varrho$, the weight matrices are initialized according to the Xavier initialization \cite{kumar_weight_2017} scheme, and the biases are initialized to zero values.

% normalization
We normalize the data set so that the spatial coordinates, the field values, and the times have zero mean and unit variance.
% batchsize
Our training pipeline disaggregates simulation snapshots into tuples $(\vect{x}, t, \vect{\mu}, \vect{u}(\vect{x}, \cdot))$ and stochastically composes the dataset $\mathcal{D}$ into $100$ equally-sized mini-batches.
% optimizer
For SNFL-ROM, we use the Adam optimizer \cite{kingma_adam_2017} for learning $\theta$, and $\varrho$.
For SNFW-ROM, we use the AdamW optimizer \cite{loshchilov_decoupled_2017} is employed as described in \autoref{subsubsec:weight}.
While we formulate weight regularization as an additional loss term, it can be easily implemented by modifying the weight-decay approach in an AdamW optimizer.
Rather than computing the sensitivity of $L_\text{Weight}$ in \autoref{eqn:weight-decay} with respect to $\theta$ via backpropagation, the expressions can be written in closed form as
\eqn{
    \label{eqn:weight-decay-grad}
    \nabla_{W^{ij}_l}L_\text{Weight} = \gamma W^{ij}_l, \hspace{1em} 
    \nabla_{\theta_i} L_\text{Weight} = 0 \hspace{1em} \text{for all other } \theta_i.
}
The sensitivity $\nabla_\theta L_\text{W}$ is computed per \autoref{eqn:weight-decay} and subtracted from $\theta$ at every optimization step \cite{loshchilov_decoupled_2017}.

% LR
We adopt the learning rate decay strategy outlined in \cite{chen_crom_2023}.
We train for $200$ epochs at each learning rate in the schedule,
\eqn{
    \label{eqn:lr_decay}
    100\cdot \eta \to
    10 \cdot \eta \to
    5  \cdot \eta \to
    1  \cdot \eta \to
    0.5\cdot \eta \to
    0.2\cdot \eta \to
    0.1\cdot \eta,
}
with a base learning rate of $\eta = 10^{-4}$.
For the 2D Burgers problem, we reduce the number of epochs from $200$ to $30$ per learning rate.
This is done to reduce training time.
Additionally, we run a learning rate warmup cycle for $10$ epochs at $\eta = 0.01$.
Early stopping is imposed if the loss value does not decrease for $20\%$ of consecutive epochs at any learning rate.
When training is complete, we find that all models have $\bar{L}_\text{data}(\theta, \varrho; \mathcal{D}_\text{train}) \sim 10^{-6}$ or lower.

%%%%%%%%%%%%%%%%%%%%%%%%%%%%%%%%%%%%%%%%%%%%%%%%%%%%%%%%%%%%%%%%%%%

%%%%%%%%%%%%%%%%%%%%%%%%%%%%%%%%%%%%%%%%%%%%%%%%%%%%%%%%%%%%%%%%%%%
% bib


%\bibliographystyle{elsarticle-num} 
%\bibliographystyle{mybibstyle} 
\bibliographystyle{unsrt} 
\bibliography{ML-ROM}





% \end{thebibliography}
%%%%%%%%%%%%%%%%%%%%%%%%%%%%%%%%%%%%%%%%%%%%%%%%%%%%%%%%%%%%%%%%%%%
\end{document}